\documentstyle[12pt,aaspp4]{article}

\def\etal{{\it et al.~\/}}

\def\kms{km s$^{-1}$}
\def\eg{{\it e.g.}}

\def\Mdot{\dot M}
\def\lta{\lower.5ex\hbox{\ltsima}}
\def\gta{\lower.5ex\hbox{\gtsima}}
\def\ltsima{$\; \buildrel < \over \sim \;$}
\def\lsim{\lower.5ex\hbox{\ltsima}}
\def\gtsima{$\; \buildrel > \over \sim \;$}
\def\gsim{\lower.5ex\hbox{\gtsima}}

\lefthead{Wilkin \& Stahler}
\righthead{Protostellar Wind Breakout}

\received{2002 December 20}
\begin{document}

\title{Trapped Protostellar Winds and their Breakout}

\author{Francis P. Wilkin}
\affil{Instituto de Astronom\'{\i}a, UNAM, Morelia, M\'exico;\ f.wilkin@astrosmo.unam.mx}
\author{Steven W. Stahler}
\affil{Astronomy Department, University of California, Berkeley, CA 94720-3411; stahler@astron.berkeley.edu}


\begin{abstract}

Observations show that high-velocity jets stem 
from deeply embedded young stars,
which may still be experiencing infall from their parent cloud cores. Yet
theory predicts that, early in this buildup, any outgoing wind is trapped by
incoming material of low angular momentum. As collapse continues and brings in
more rapidly rotating gas, the wind can eventually break out. Here we model
this transition by following the motion of the shocked shell created by impact
of the wind and a rotating, collapsing envelope.
We first demonstrate, both analytically and numerically, that our 
previous, quasi-static solutions are dynamically unstable. Our 
present, fully time-dependent calculations  include cases both where 
the wind is driven back by infall to the stellar surface, 
and where it erupts as a true outflow.
For the latter, 
we find that the time of breakout is  $5\times 10^4\,{\rm yr}$ 
for wind speeds of $200\,{\rm km}\,{\rm s}^{-1}$.
The reason for the delay 
is that the shocked material, including the swept-up infall, 
must be able to climb out of the star's gravitational potential well. 

We explore the critical wind speed necessary for breakout 
as a function of the mass transport rates in the wind and infall, 
as well as the cloud rotation rate $\Omega_\circ$
and time since the start of infall. 
Breakout does occur for realistic parameter choices. The actual breakout 
times would change  if we relaxed the assumption of perfect mixing between 
the wind and infall material. 
Our expanding shells do not exhibit the
collimation of observed jets, but continue to expand laterally. 
To halt this expansion, the density in the envelope 
must fall off less steeply than in our model. 

\end{abstract}



%



\section{Introduction}

The traditional picture of low-mass star formation in isolation (Shu, Adams,
\& Lizano 1987) envisions an early period of pure accretion, prior to the
appearance of any wind. This expectation is based on simple theoretical
considerations. Since the star is building up mass from its parent cloud,
the rate of mass infall exceeds that of any outflow. In addition, both theory
and observation suggest that the {\it speeds} of infalling and wind gas are
comparable. Thus, the ram presure from the collapsing envelope should crush
the wind, preventing its escape from the stellar or inner disk surface. Why
is it, then, that the most embedded stars, including those designated Class 0
(Andr\'e \etal 1993) are observed to produce vigorous winds? How do wind and
infall simultaneously occur in a very young object?

The answer, as has long been appreciated, is that neither flow is isotropic.
Rotation and magnetic fields, in both the star and cloud, break spherical
symmetry and alter the geometry of the wind and infall. Moreover, rotational
distortion increases as the collapse of the parent cloud proceeds in an
inside-out fashion (Shu 1977). Our goal in this paper is to 
examine numerically the interaction between wind and infall. 
Since the full situation is complex, 
it is best to proceed in stepwise fashion. We concentrate, therefore, on the 
effect of infall geometry, 
positing for simplicity a spherical wind. We also
focus exclusively on the rotational influence. That is, 
we neglect any magnetic
force on the infalling gas, under the assumption that decoupling from the
ambient field has already occurred (Li \& McKee 2000).

As in our first study of this topic (Wilkin \& Stahler 1998; 
hereafter Paper I),
we follow the dynamics of the thin shell formed by the colliding wind and
infall. Paper I derived an evolutionary sequence of shells under 
the simplifying
assumption of quasi-steady flow. That is, 
we took the expansion or contraction
of the shell to be slow compared to either the wind or infall speeds. We now
study the matter in more detail, solving the initial value problem of
shell motion in a fully time-dependent manner. Our finding, in brief, is that 
the shell can either break out or recollapse to the star, depending on both 
the flow parameters and the epoch during cloud collapse. We derive the time 
of breakout as a function of the wind and infall parameters, 
and explain how to generalize this to the case of an anisotropic 
driving wind. 

In \S2 below, we first detail the mathematical formulation of the 
problem, obtaining the time-dependent version of the dynamical equations. 
After
describing our method of solution in \S 3, we show in \S 4 that our
previous, quasi-steady shells are in fact dynamically unstable. We set up our
initial conditions and nondimensionalization of the launch problem in \S
5, before presenting an overview of our numerical results in \S 6. We also
rederive key findings in a heuristic, analytical fashion. Finally, Section 7
discusses the broader astrophysical implications of our study.
In particular, an  observational signature of the trapped wind phase 
may be fluctuations of the accretion luminosity, 
as an oscillatory shock structure is  expected to develop prior 
to breakout of the wind.

\section{Formulation of the Problem}
\subsection{Thin-Shell Approximation}

The supersonic collision of wind and infall leads to the formation of both
an inner and outer shock front. At the wind speeds appropriate for low-mass
stars (typically less than $300\, {\rm km}\, {\rm s}^{-1}$), the cooling behind these shocks
is relatively efficient. (Recall the discussion in \S 2.2 of Paper I.) Hence
we may describe the intershock region as a cold, thin shell, 
characterized by its
radius $R(\theta ,t)$ and mass surface density, $\sigma (\theta ,t)$. Here,
$\theta$ is the polar angle measured from the rotation axis of the parent 
cloud. As in Paper I, we assume this shell to be axisymmetric, i.e., invariant
with respect to the azimuthal angle $\phi$.

Within the shell, two fluids with very different temperature and velocity come
into contact. The internal shearing layers are subject to the Kelvin-Helmholtz
instability, which leads quickly to turbulent mixing. We shall assume that the
mixing is so efficient that we may describe the shell as a single fluid with
a time-averaged velocity at each point $(R, \theta, t)$. This velocity has an
azimuthal component arising from rotation of the infalling envelope. It also
has a meridional component along the shell. Furthermore, we do {\it not}
neglect, as in Paper I, the velocity of the shell normal to the surface.

The evolution of the shell is governed by the incident fluxes of mass and
momentum from the wind and infall. It is also influenced critically by the
gravitational attraction of the star for the shell material. The central 
difference of the present study and Paper I is that we no longer assume the
evolution to be quasi-steady. 
That is, the properties of the shell are allowed to change 
over the time required for material to travel along the surface.

\subsection{Description of Infall and Outflow}

The infall arises from the gravitational collapse of a dense
cloud core within a larger molecular cloud. 
To obtain the resulting accretion flow, we 
idealize this core as a singular isothermal sphere rotating rigidly with 
angular velocity $\Omega_\circ$. 
Our calculation does not address the
formation of the core itself, and neglects any turbulence in the initial 
structure.  Recent theoretical work (e.g., Cho, Lazarian, \& Vishniac 2002; 
Vazquez-Semadeni, Ballesteros-Paredes, \& Klessen 2003) 
suggests how cores might  condense out of a turbulent flow. 
Nevertheless,  observations continue to show that nonthermal motion 
in the core itself is  relatively small (Barranco \& Goodman 1998). 
Once the infalling gas becomes supersonic, the effects of this 
turbulent component  are expected to be minor. 
Our adoption of a spherical body, while clearly an  idealization, 
should be of little consequence for  the  flow deep within the core's center.

In any event, the distributions of infalling density $\rho_i$ 
and velocity ${\bf u}_i$ depend on time, because the collapsing region 
spreads out as a rarefaction wave at the 
isothermal sound speed $a_\circ$,  gradually engulfing material 
of higher specific angular momentum (Shu 1977).
We adopt the infall model of Cassen \& Moosman (1981), which represents 
the inner limit of the full collapse. 
This limit applies to  radii  that are small compared to 
the expansion front $R_{exp} = a_\circ\,t$, 
where $t$ is the time since the collapse began. 

We may gauge the total rate of infall by 
examining the transport of mass across a surface located well inside 
the expansion front. According to the inside-out collapse model of 
Shu (1977), this transport rate is:
\begin{equation} \Mdot_i = {{m_\circ\,a_\circ^3}\over G} \,\,,\end{equation}
where $m_\circ = 0.975$. 
The infall itself is characterized by a time-dependent length scale, the 
centrifugal radius $R_{\rm cen}$. This is given by
\begin{eqnarray}
 R_{cen} & = &\frac{1}{16}\,
m_\circ^3\,a_\circ\,\Omega_\circ^2\,t^3,\nonumber\\
R_{cen}({\rm AU})& = & 
{{a_\circ} \over {0.2\,{\rm km}\,{\rm s}^{-1}}}
\biggl({{\Omega_\circ} \over {2\times 10^{-14}\,{\rm s}^{-1}}}\biggr)^2
\biggl({{t} \over {10^5\,{\rm yr}}}\biggr)^3.
\end{eqnarray}
Our reference value for $\Omega_\circ$ is chosen to be consistent
with inferred rotation rates based upon velocity gradients in cloud
cores lacking embedded IRAS sources (Jijina, Myers, \& Adams 1999). 
For an equator-on rotating core, this
corresponds to a velocity gradient of 
$0.6\,{\rm km}\,{\rm s}^{-1}\,{\rm pc}^{-1}$. 
In regions where $r \lsim R_{cen}$, 
rotational distortion of the flow is significant, 
and some of the infalling matter strikes the equatorial plane 
($\theta = \pi/2$), forming a circumstellar disk.  
When $r >> R_{cen}$, the infall is nearly spherically 
symmetric. Equation (1) gives the {\it sum} of the mass per 
unit time impacting the star directly and that entering the disk.

In terms of the nondimensional radial variable $\zeta \equiv R_{cen}/r$, 
the infall  velocity components and density are (Terebey, Shu \& Cassen 1984)
\begin{eqnarray}
u_{i,r} & = & -\left({{GM_{*}}\over r} \right)^{1/2}
\left(1+{{{\rm cos}\,\theta}\over{{\rm cos}\,\theta_\circ}}\right)^{1/2}\\
u_{i,\theta} & = & \left({{G M_{*}}\over r}\right)^{1/2}
\left({{{\rm cos}\,\theta_\circ-{\rm cos}\,\theta}\over{{\rm sin}\,\theta}}
\right)
\left(1+{{{\rm cos}\,\theta}\over{{\rm cos}\,\theta_\circ}}\right)^{1/2}\\
u_{i,\phi} & = & \left({{G M_{*}}\over r}\right)^{1/2}
{{{\rm sin}\,\theta_\circ}\over{{\rm sin}\,\theta}}
\left(1-{{{\rm cos}\,\theta}\over{{\rm cos}\,\theta_\circ}}\right)^{1/2}\\
\rho_i & = & -{{\Mdot_i}\over{4\pi r^2 u_{i,r}}}
\left[1+2\,\zeta\,P_2\left({\rm cos}\,\theta_\circ\right)\right]^{-1}\,\,. 
\end{eqnarray}
The function $P_2({\rm cos}\,\theta_\circ)$ is the Legendre 
polynomial, where $\theta_\circ$ is a Lagrangian variable that labels the 
streamlines. Specifically, $\theta_\circ$ is the initial 
polar angle of each 
fluid element, just before it is overtaken by the rarefaction wave. 
This angle is given implicitly in terms of the instantaneous coordinates 
$\zeta$ and $\theta$ by the trajectory equation 
\begin{equation} 
\zeta = {{{\rm cos}\,\theta_\circ - {\rm cos}\,\theta}
\over
{{\rm sin}^2\,\theta_\circ\,{\rm cos}\,\theta_\circ}}
\,\,.
\end{equation}

Our goal is to assess the collimating influence of the anisotropic
infall. Accordingly,  we idealize the wind to 
be spherically symmetric, with an associated mass transport rate of 
${\dot M}_w$. 
If the star lies at the origin of a spherical coordinate system, 
then the wind density at a radial distance $r$ is
\begin{equation} 
\rho_w  =  {{\Mdot_w}\over{4\,\pi\,r^2\,V_w}}.
\end{equation}
Here, $V_w \equiv u_{w,r}$ is the wind velocity. Note that the 
components $u_{w,\theta}$ and $u_{w,\phi}$ are both zero.

We do not investigate the structure of the star, which is taken simply 
to be a spherical, gravitating mass. The value of this mass is 
\begin{equation} 
 M_{*} = {\dot M}_i \, t\,.
\end{equation} 
Thus, we assume that both the matter entering the disk and that 
colliding with the infall are efficiently recycled to the star. How  
gas within the shocked shell reaches the disk is, of course, one of  
our main concerns. On the other hand, the transport of matter 
{\it within} the disk through internal torques is beyond the scope 
of this project. 
Once the shell truly moves
dynamically, mass accretion onto the star and disk is halted, and we 
freeze the stellar mass at the value corresponding to the launch time 
of the shell (See \S4).  

\subsection{Time-Dependent Equations: Mathematical Derivation}


As in Paper I, we consider a small, three-dimensional patch of the shell,  
whose center is located at $(R,\theta,\phi)$ within a global, spherical  
coordinate system centered on the star. We also utilize a global, Cartesian  
system $(x,y,z)$. (See Figure 3 of Paper I, i.e., Figure I.3) 
To help in following our necessarily abbreviated 
derivation of the equations, we advise the reader to consult \S 2.3 of 
Paper I.

Figure~\ref{fig:wsIf4}, a slightly modified version of Figure I.4, shows the representative
patch in more detail. We let $\gamma$ denote the surface tilt, i.e., 
the angle 
between the patch normal ${\hat {\bf n}}$ and the radial direction
from the star ${\hat {\bf r}}$.  The upper and lower faces, i.e., those 
depicted with the largest areas, coincide with 
the inner and outer shock fronts. The narrow faces
to the left and right of the upper one trace loci of constant polar angle
$\theta$, while the remaining two have fixed $\phi$. Note that the length
$\Delta s$ is a small increment of the global coordinate $s$ measuring 
distance along the shell from the pole to the equatorial plane. Inspection of
Figure~\ref{fig:wsIf4} shows that $\Delta s = R\,{\rm sec}\,\gamma\,\Delta\,\theta $ 
and that the patch width is 
$\Delta w = \,R\,{\rm sin}\,\theta\,\Delta\,\phi$. 
The surface area of either the upper or lower face is 
$\Delta s \Delta w = {\cal A} \Delta \theta \Delta \phi$, 
and the total patch volume is $\Delta s \Delta w \Delta n = 
{\cal A}\Delta\theta\Delta\phi\Delta n$.
Here $\Delta n$ (called $\Delta h$ in Paper I) is the patch thickness. 
The factor ${\cal A}$ is given by
\begin{equation} 
{\cal A} = R^2 {\rm sin}\theta\, {\rm sec}\gamma.
\end{equation} 

For a mathematical description of the shell evolution, we let 
$R = R(\theta, \phi, t)$ be a dependent variable. 
Then the tilt angle $\gamma$ 
is given by  $\tan\,\gamma = -R'/R$, 
where $R' \equiv \partial R/\partial\theta$. 
During the evolution, we want the sides of our patch to retain fixed angular
positions in $\theta$ and $\phi$. On the other hand, we allow the upper and
lower surfaces to move in and out radially, in order to follow expansion or
contraction of the shell. Let us denote by ($u_r, u_\theta, u_\phi$) the
velocity of matter within the shell. (Note the lack of additional subscripts, 
which we use to denote wind or infall.) 
The velocity component of this fluid pointing 
along the patch normal is
\begin{equation} 
 u_{n}\,  =  \,u_{r}\,{\rm cos}\,\gamma\,+
            \,u_{\theta}\,{\rm sin}\,\gamma.
\end{equation} 
Now shell expansion gives the patch itself a normal velocity
\begin{equation} 
v_n = {{\partial \,R} \over {\partial\,t}} \cos\gamma.
\end{equation} 
Our kinematical constraint on the patch motion is simply $u_n = v_n$. 
Combining equations (11) and (12), we find
\begin{equation} 
{{\partial R}\over {\partial\, t}} = 
u_r\, +\, u_\theta\,\tan\gamma.
\end{equation} 

Proceeding to the dynamical equations, we first derive an expression for
$\Delta{\dot Q}$, the rate of change of any physical quantity within 
the patch.
Let $q$ denote the volume density of this quantity. This density may change 
with time, as may the patch volume. We thus have
\begin{equation} 
\Delta {\dot Q} = {{\partial} \over {\partial t}}
 \biggl(q\, \Delta\,n\,\Delta\,s\, 
\Delta w\biggr) = {{\partial} \over {\partial t}} \,\biggl(
{\cal A}\, q\, \Delta n\biggr)\,\Delta\,\theta\,\Delta\,\phi\,.
\end{equation} 
where the partial derivative is at fixed $\theta$ and $\phi$.

The change in our quantity comes in part from advection into and
out of the patch. The rate here depends on both the wind and 
infall velocities,
and those within the patch itself. Note that the relevant wind and infall
velocities are the normal components {\it relative to the patch}, since the 
latter has the normal velocity $u_n$. Let $q_w$ and $q_i$ 
be the density of the
quantity in the wind and infall, respectively. Then these flows make an
{\it external} contribution to $\Delta{\dot Q}$:
\begin{eqnarray}
 {\Delta{\dot Q}}_{\rm ext} &  = & 
\big[(u_{w,n}-u_n)\,q_w\,-\,(u_{i,n}-u_n)\,q_i]\,\Delta s\,\Delta w,\nonumber\\
& = & \,{\cal A}\,\bigl[{u}_{w,n}'\,q_w\, -\, {u}_{i,n}'\,q_i\bigr]\,
\Delta\,\theta\,\Delta\,\phi\,\,.\end{eqnarray}
where we have used primes for the relative normal velocities.

We must also account the {\it internal} contribution to advection, i.e., 
the effect of flow within the shell. Part of
this flow is due to motion in the $\theta$-direction. Noting that the
radial thickness of the shell is $\Delta r = (\Delta n) \,\sec\gamma$, 
Figure~\ref{fig:wsIf4} shows that
\begin{eqnarray}
{\Delta{\dot Q}}_{\rm int,\theta} & = &   
[u_\theta\,q\,\Delta r\,\Delta w]_{\theta-\Delta\theta/2}\,
-\,  [u_\theta\,q\,\Delta r\,\Delta w]_{\theta+\Delta\theta/2}\,\nonumber\\
 &  = & -{{\partial}\over{\partial\,\theta}}\,
\biggl({\cal A}\,{{u_\theta\,q\,\Delta\, n} \over {R}}\biggr)\,
\Delta\,\theta\,\Delta\,\phi\,\,.\end{eqnarray} 
The final contribution to advection stems from the $\phi$-velocity. 
Here we find\hfil\break
\begin{eqnarray}
{\Delta{\dot Q}}_{\rm int,\phi} \, & = & \,
[u_\phi\,q\,\Delta\,s\,\Delta\,n]_{\phi-\Delta\phi/2} - 
[u_\phi\,q\,\Delta\,s\,\Delta\,n]_{\phi+\Delta\phi/2} 
\nonumber\\
 & = & \, - u_\phi\,R\,\sec\gamma\,
{{\partial q} \over {\partial \phi}}\,\Delta\,n\,\Delta\,\theta\,\Delta\,\phi.
\end{eqnarray}
Apart from notation, this equation is identical to equation (I.18).

We now sum all the advective terms to form $\Delta{\dot Q}_{\rm adv}$.
Combining equations (15)-(17), we have
\begin{eqnarray}
 \Delta\, {\dot Q}_{adv} &=& 
\biggl\{ - {{\partial} \over {\partial \theta}} 
\biggl({\cal A}{{u_\theta} \over R} q\,\Delta n \biggr)
 +  {\cal A} \biggl[u_{w,n}' q_w 
- u_{i,n}' q_i\nonumber\\
 & -& {{u_\phi} \over {\varpi}} {{\partial} \over {\partial \phi}}
(q\, \Delta n)\biggr]\biggr\}\,\Delta\,\theta\,\Delta\,\phi\,,
\end{eqnarray} 
where $\varpi \equiv R\,\sin\theta$ is the cylindrical radius. 

In order to express mass conservation, we let $q$ be the mass density $\rho$.
Note that this quantity, like the others we shall be considering, is
formally infinite in the limit $\Delta n \rightarrow 0$, 
whereas $\rho\, \Delta n$ is the
finite surface density $\sigma$. Since there are no sources or sinks of mass,
we demand that $\Delta{\dot Q} = \Delta{\dot Q}_{\rm adv}$. We find
\begin{eqnarray}
{{\partial} \over {\partial t}}\biggl({\cal A}\,\sigma\biggr) + 
{{\partial} \over {\partial \theta}}
\biggl({\cal A}{{\sigma\, u_\theta}\over R}\biggr) 
 = {\cal A}\, \biggl\{\rho_w u_{w,n}' - \rho_i u_{i,n}'\biggr\}.
\end{eqnarray}

\indent
Turning to momentum, we first note that the magnitude of the 
gravitational force on the patch is 
$\Delta \,F_g = GM_{*}\sigma{\cal A}\Delta\theta\Delta\phi/R^2$. 
This force would increase the momentum within the patch even if
there were no advection. 
Suppose we choose our patch to be located at $\phi = 0$ in the global, 
spherical coordinate system. Then the Cartesian components 
of the gravitational force become
$\Delta \,F_{g,x} = - \Delta F_g\,\sin\theta$, 
$\Delta F_{g,y} = 0$, and $\Delta \,F_{g,z}= - \Delta \,F_g\cos\theta$. 
The physical conservation law is most directly expressed
in terms of Cartesian coordinates.
In the $x$-momentum equation, we let $q = \rho u_x$ in equations (14) and 
(18), and
demand that $\Delta\,{\dot Q}\,=\,\Delta{\dot Q}_{adv}\,+\,\Delta F_{g,x}$. 
For the patch centered on
$\phi = 0$, $u_x \approx u_\varpi$ and 
$\partial u_x/\partial\phi \approx -u_\phi$. 
Here $u_\varpi$ is the cylindrical radial component of velocity.
We obtain
\begin{eqnarray}
{{\partial} \over {\partial\, t}}\biggl({\cal A}\,\sigma\,
u_\varpi\biggr) + {{\partial} \over {\partial\, \theta}}\biggl({\cal A}\,
{{\sigma\, u_\theta\, u_\varpi} \over {R}}\biggr) = {\cal A}\,\biggl\{ 
\rho_w\, u_{w,n}'\, u_{w,\varpi}\nonumber\\
 - \rho_i\, u_{i,n}'\, u_{i,\varpi}\, 
+ {{\sigma\, u_\phi^2} \over {\varpi}} 
- {{G\, M_{*}\,\sigma} \over {R^2}}\,\sin\,\theta\biggr\}.
\end{eqnarray}
The $y$-component of momentum conservation is simpler 
since $\Delta \,F_{g,y}=0$.  
We now let $q = \rho\,u_y$, and use $u_y \approx u_\phi$ and
$\partial\, u_y\,/\,\partial\,\phi \approx u_\varpi$ to find
\begin{eqnarray}
  {{\partial} \over {\partial\, t}}
\biggl({\cal A}\,\sigma\, u_\phi\biggr)
+  {{\partial} \over {\partial\, \theta}}
\biggl({\cal A}\,{{\sigma\, u_\theta\, u_\phi} \over R}\biggr)
=  {\cal A}\,
\biggl\{\rho_w\, u_{w,n}'\, u_{w,\phi}\nonumber\\
 - \rho_i\, u_{i,n}'\, u_{i,\phi}
 - {{\sigma u_\phi u_\varpi} \over {\varpi}}\biggr\}.
\end{eqnarray}
We may obtain a simpler form of this $\phi$-force equation 
by instead writing it in terms of the z-component of angular momentum. 
First note that using equation (13), we have the relation 
$\partial\varpi/\partial\,t + (u_\theta/R)\partial\varpi/\partial\theta = u_\varpi$.
Multiplying this equation by ${\cal A}\sigma u_\phi$, and adding it to
$\varpi$ times equation (21), we obtain 
\begin{eqnarray}
{{\partial} \over {\partial\, t}}
\biggl({\cal A}\,\varpi\, \sigma\, u_\phi\biggr) 
+ {{\partial} \over {\partial\, \theta}}
\biggl(\varpi{\cal A}\,{{\sigma\, u_\theta\, u_\phi} \over R}\biggr) = 
  {\cal A}\,\varpi\,\biggl\{\rho_w\, u_{w,n}'\, u_{w,\phi}\nonumber\\ 
- \rho_i\, u_{i,n}'\, u_{i,\phi}\biggr\}.
\end{eqnarray}
Similarly, the z-force equation is obtained using $q = u_z$:
\begin{eqnarray}
{{\partial} \over {\partial\, t}}\biggl({\cal A}\,\sigma\,
u_z\biggr) + {{\partial} \over {\partial\, \theta}}\biggl({\cal A}\,
{{\sigma\, u_\theta\, u_z} \over {R}}\biggr) = {\cal A}\,\biggl\{ 
\rho_w \,u_{w,n}'\, u_{w,z}\nonumber\\ - \rho_i\, u_{i,n}'\, u_{i,z} 
 - {{G\, M_{*}\,\sigma} \over {R^2}}\,\cos\,\theta\biggr\}.
\end{eqnarray}

We have seen that beginning with Cartesian components of velocity, the
azimuthal symmetry of the problem leads naturally to equations in terms
of the cylindrical components of velocity. However, for a radial driving wind,
it will be most useful to use spherical components. 
By taking linear combinations of equations (20) and (23), 
and substituting $u_\varpi = u_r\sin\theta+u_\theta\cos\theta$ and 
$u_z = u_r\cos\theta-u_\theta\sin\theta$, we may recast
them in terms of the velocities $u_r$ and $u_\theta$:
\begin{eqnarray}
{{\partial} \over {\partial\, t}}\biggl({\cal A}\,\sigma\,
u_r\biggr) + {{\partial} \over {\partial\, \theta}}\biggl({\cal A}\,
{{\sigma\, u_\theta\, u_r} \over {R}}\biggr) & = & {\cal A}\,\biggl\{ 
\rho_w\, u_{w,n}'\, u_{w,r}\nonumber\\ - \rho_i\, u_{i,n}'\, u_{i,r}\, 
+ {{\sigma\, u_\phi^2} \over {R}} 
- {{G\, M_{*}\,\sigma} \over {R^2}}\,\biggr\},\\
{{\partial} \over {\partial\, t}}\biggl({\cal A}\,\sigma\,
u_\theta\biggr) + {{\partial} \over {\partial\, \theta}}\biggl({\cal A}\,
{{\sigma\, u_\theta\, u_\theta} \over {R}}\biggr) &  = & {\cal A}\,\biggl\{ 
\rho_w \,u_{w,n}'\, u_{w,\theta}\nonumber\\
- \rho_i\, u_{i,n}'\, u_{i,\theta}
+ {{\sigma\, u_\phi^2\cot\theta} \over {R}} \biggr\}.\end{eqnarray}
Previously, Giuliani (1982) derived an equivalent set of 
equations, including magnetic fields, 
but without accounting for gravity or rotation.

\subsection {Time-Dependent Equations: Lagrangian Form}

In practice, it is simplest to solve the evolutionary equations by 
recasting them in Lagrangian form. 
Consider first the comoving derivative of any quantity $Q$ within the shell. 
In the Eulerian description we have used until now, $Q$ is a function of
$t$, $\theta$, and $\phi$. It also depends implicitly on radius, since, at
fixed $\theta$ and $\phi$, the radius $R$ varies with time. The comoving
derivative is therefore
\begin{equation}
 {{DQ} \over  {D\,t}} = {{\partial\,Q} \over {\partial t}} 
+  {{u_\theta} \over{R}}{{\partial\,Q} \over {\partial \theta}}
+ {{u_\phi}\over {R\,\sin\theta}}
{{\partial\,Q} \over {\partial \phi}}.\end{equation}
We now let $Q$ be, in turn, the radial and angular positions of a moving
fluid element that constitutes part of the shell. This element is no longer constrained in $\theta$ and $\phi$, as was our patch. 
Retaining the old notation for these
coordinates, now serving as dependent variables, we substitute into 
equation (26) to find
\begin{eqnarray}
 {{D\,R} \over {D\,t}}  & = & u_r,\\
{{D\,\theta} \over {D\,t}} & = & {{u_\theta} \over R},\\
{{D\,\phi} \over {D\,t}} & = & {{u_\phi} \over {R \sin\theta}}.
\end{eqnarray}
In deriving equation (27), we have used both equation (13) and the definition 
of $\gamma$ in terms of $R^\prime$. Equations (27)-(29) 
describe both the expansion of the shell 
and tangential motion along its surface.
We next substitute for $Q$ in equation (26) the quantity ${\cal A}\sigma$. 
After utilizing equation (19), we arrive at the Lagrangian expression for
mass conservation, which we may write as
\begin{equation}
{{D\,({\rm ln}({\cal A}\,\sigma))} \over {D\,t}}
 =  \bigl[\rho_w\,u_{w,n}' - \rho_i\,u_{i,n}'\bigr]\,/\,\sigma 
  - {{\partial} \over {\partial \theta}}\biggl({{u_\theta} \over
{R}}\biggr).
\end{equation}
To obtain the $r$- and $\theta$- components of the force equation, as well
    as the expression for angular momentum conservation, we substitute for
    $Q$ the quantities $u_r$, $u_\theta$, and $\varpi u_\phi$,
    respectively. We then use our Eulerian conservation laws and equation (30)
    in each case to find
\begin{eqnarray}
 {{D\,u_r} \over {D\,t}} & = &   \ [\rho_w\, u_{w,n}'\, u_{w,r}' 
- \rho_i\, u_{i,n}'\, u_{i,r}']/\sigma + {{{u_\theta^2}} \over R}
 + {{{\l}^2 \sin\theta} \over {{\varpi}^3}} 
- {{G\, M_{*}} \over {R^2}},\\
 {{D\,u_\theta} \over {D\,t}} & = & 
  \ [\rho_w\,u_{w,n}'\,u_{w,\theta}'\, 
- \,\rho_i\,u_{i,n}'\,u_{i,\theta}']/\sigma  - {{u_r\, u_\theta} \over R} 
 + {{{\l}^2 \cos\theta} \over {{\varpi}^3}},\\
{{D\,\l} \over {D\,t}} & = & 
[\rho_w\,u_{w,n}'\,\l_w' - \rho_i\,u_{i,n}'\,\l_i']\,/\,\sigma.
\end{eqnarray}
Here we have let $l = \varpi u_\phi$ be the $z$-component of specific angular
    momentum, and have further defined
\begin{eqnarray}
 \l_w' & = & \varpi\, (u_{w,\phi}-u_\phi),\\
 \l_i' & = & \varpi\, (u_{i,\phi}-u_\phi).
\end{eqnarray}

Equations (27)-(29) and (31)-(33) may be summarized in vector form as
\begin{eqnarray}
{D{\bf R}} \over {D\,t} &  = & {\bf u},\\
 {{D\,{\bf u}} \over {D\,t}} & =  & 
[\rho_w\, {u}_{w,n}'\, {\bf u}_w' 
 - \rho_i\, {u}_{i,n}'\, {\bf u}_i']/\sigma  
 - {{G\, M_{*}} \over {R^2}}\,{\hat {\bf r}}.\end{eqnarray}
As expected, the acceleration of the fluid element 
depends on both the 
radial force of gravity and on the input of momentum 
from wind and infall. Note finally that the mass conservation equation (30) 
contains an Eulerian derivative on the righthand side, so that our equations 
are not in purely Lagrangian form. Indeed, the normal components of wind and 
infall speed that enter equations (30)-(33) also require, through the 
angle $\gamma$, the Eulerian derivative $\partial R/\partial\theta$. In 
practice, this mixed character of the equations, and the fact that 
$\theta$ acts as both a dependent and independent variable, present no 
special difficulty for integration.

\section{Method of Solution}



    Equations (27)-(33) fully describe the motion of a fluid element within 
    the shell. Because the latter is axisymmetric, there is no need to track 
    the $\phi$ coordinate, and we may ignore equation (29). We treat the 
    others effectively as ordinary differential equations in time, and use 
    them to follow a collection of 50 discrete points 
    that represents our shell. 
    We evaluate the cross derivatives $\partial R/\partial\theta$ and 
    $\partial(u_\theta/R)/\partial\theta$, 
by numerically differencing $R$ and $u_\theta/R$. 
This method has been employed previously for 
the similar problem of steady-state, 
non-axisymmetric bow shocks and expanding supershells 
(Mac Low and McCray 1988; Bandiera 1993), 
and works well provided the spacing between adjacent trajectories
is fine enough to determine numerical derivatives.

    For each of the 50 points, we integrated the equations in time using a
    fifth-order Runge-Kutta scheme. Cross derivatives were obtained at
    each point by using the two nearest neighbors, and were thus
    second-order accurate. Since the points move in a manner dictated by the
    local fluid velocity, they are unevenly spaced in angle. In practice, we
    approximated the shell segment by fitting a circular arc through each
    triad of points. The cross derivatives at the middle point were then
    obtained analytically from this curve.

    As the shell evolved in time, some of its representative points inevitably
    drifted toward the equatorial plane. In addition, points could move so 
    close together that there was little variation in the intervening space.
    In either circumstance, we removed a point from the original set. We
    immediately replaced it with another, which we introduced at the part of
    the shell that was most sparsely covered. We always retained one point on
    the symmetry axis, where the boundary conditions $R'=u_\theta=u_\phi=0$ 
were applied. Here the mass equation was written in modified form to account 
for the  vanishing of the $\sin\theta$ factor within  ${\cal A}$.

We tested our code on three problems with known, analytic solutions. 
They include: (1) an expanding, spherical shell driven by an  
isotropic wind, (2) a shell driven 
by an angle-dependent wind within an $r^{-2}$ density distribution 
(Shu \etal 1991), 
and (3) the bowshock created by the wind from a star that is 
moving into a uniform medium. 
The first two tests confirmed the code to better than $1\%$ accuracy
in a non-steady situation. 
For the  last problem, Wilkin (1996) found analytic 
solutions for the steady-state 
shell configuration, including the surface density 
and flow velocity along the shell. 
Figure~\ref{fig:wsIf3} shows the expansion of our time-dependent bowshock 
as it approaches the steady-state endpoint. 
We have checked  that the approach to
equilibrium agrees with the calculations of Giuliani (1982).
Figure~\ref{fig:errors}  compares a number of quantities 
in the analytic solution with our numerical results, 
at a time when equilibrium has been reached. The fractional 
errors are between $10^{-2}$ and $10^{-4}$, except close to the pole. 
Judging from the three tests, we believe our code is 
capable of tracking all variables in the time-dependent shell to within 
an accuracy of 1-2 percent.

\section{Instability of the Quasi-Steady Solution}

Before displaying our fully time-dependent results, we first discuss the 
quasi-steady solutions obtained in Paper I. 
\footnote{We draw the reader's attention to two typographical errors
in Paper I. Equation (70) should have an overall minus sign on the right 
hand side, 
and  equation (72) should read 
$ \epsilon\,\equiv\,\alpha\,{\rm sin}\,\theta\,-\,
\tau\,\Phi_m^2/R\Phi_t\,{\rm cos}\,\gamma\,\,.$}
These shells are in dynamical equilibrium,
with gravity 
and the infall ram  pressure opposing the outward ram pressure from the wind. 
Because gravity plays a dominant role, we suspected that the shells might be 
dynamically unstable. (See \S 4 of Paper I.) We now demonstrate this fact 
both analytically and numerically.

\subsection{Analytical Argument}

We first examine the stability in the region of the polar axis. This
approach extracts the basic physics while bypassing a full modal analysis,
such as that done for wind bowshocks by Dgani, Van Buren, \& Noriega-Crespo
(1996). The total radial force per unit area acting on the shell near the
axis is
\begin{equation}
 F_r = \rho_w V_w^2 - \rho_i u_{i,r}^2 - {{G M_{*} \sigma} 
\over {R^2}},\end{equation}
   Note that we have omitted centrifugal terms, which are vanishingly small
   near the pole.  

   The condition of normal force balance is that $F_r (R_\circ) = 0$, where
   $R_\circ$ is the shell's equilibrium polar radius. We now imagine 
   perturbing the shell, and examine the resulting change in $F_r$. For this
   purpose, we consider the lowest-order oscillation mode of the shell, i.e.,
   its breathing mode. We find the altered wind density $\rho_w$ from 
   equation (8), and the infall density $\rho_i$ and velocity $u_{i,r}$ from
   equations (6) and (3), respectively. For the latter two quantities, we
   specialize to the pole by setting $\theta_\circ = \theta = 0$. To obtain
   the perturbed value of the mass density $\sigma$, we ignore any additional
   mass swept up by the shell during its oscillation. Thus, to conserve mass, 
     we write
   $\sigma = \sigma_\circ R_\circ^2/R^2$, where $\sigma_\circ$ is the 
   equilibrium value. Equation (38) then becomes
\begin{equation}
F_r = {{{\dot M}_w V_w} \over {4 \pi R^2}} 
- {{{\dot M}_i} \over {4 \pi R^2}} 
\biggl({{2 \,G M_{*}} \over {R}}\biggr)^{1/2} {1\over {1+2 \zeta}} 
- {{G M_{*}\,\sigma_\circ\,R_\circ^2} \over {R^4}}.
\end{equation}
  where $\zeta \equiv R_{\rm cen}/R$. Note that we have also neglected
   alterations to the surface density due to {\it the perturbation} in the 
tangential motion. We showed in Paper I 
that such motion is generally small compared to the wind and infall speeds.

We next nondimensionalize equation (39) by dividing through by 
${\dot M}_w V_w/4\pi R_\circ^2$. 
Again employing subscripts for equilibrium values, we have 
\begin{equation}
 {\cal F}_r = \biggl({R\over {R_\circ}}\biggr)^{-2} 
- f_i \biggl({R\over {R_\circ}}\biggr)^{-5/2} 
{{1+2\, \zeta_\circ}\over {1+2\, \zeta}} 
- f_g \biggl({R\over {R_\circ}}\biggr)^{-4}.\end{equation}
  The constants $f_i$ and $f_g$ are the nondimensional (fractional) 
contributions 
of the  infall and gravitational forces, respectively, at the 
equilibrium position.   These are given by
\begin{eqnarray}
 f_i & = & {{{\dot M}_i} \over {{\dot M}_w}} \biggl({{2 \,G M_{*}} 
\over {V_w^2 R_\circ}}\biggr)^{1/2} {1\over {1+2 \zeta_\circ}},\\
 f_g & = & {{4\,\pi\,G M_{*}\,\sigma_\circ} \over {{\dot M}_w V_w}}.
\end{eqnarray}

The requirement that $R_\circ$ be an equilibrium radius means that 
   $f_g = 1 - f_i$. Thus, we may eliminate $f_g$ from equation (40). Writing 
   $R/R_\circ = 1 + \delta$ and recalling that $\zeta \propto 1/R$, 
   we linearize 
   in $\delta$ to find the small radial force due to the oscillation: 
\begin{equation} 
{\cal F}_r \approx \biggl[2 - f_i {{3 + 10\, \zeta_\circ} \over 
{2 (1 + 2\, \zeta_\circ)}}\biggr] \delta.\end{equation}
   We see that there is a critical value for $f_i$: 
\begin{equation}
 f_{i,crit} = {{4 (1 + 2\, \zeta_\circ)} \over {3 + 10\, \zeta_\circ}}.
\end{equation}
   If the actual $f_i$ in the equilibrium solution exceeds $f_{i, crit}$, then 
   the radial force is directed opposite to the displacement, i.e., the shell 
   is stable (or overstable). Conversely, $f_i < f_{i, crit}$ implies dynamic instability.  
The {\it minimum} value of $f_{i, crit}$, that for infinite $\zeta_\circ$, 
is 0.8. That is, at least $80$ percent of the confining force must come 
from the infall ram pressure, rather than gravity, to ensure stability.

   We now return to the exact results of Paper I. There we found $f_i$ to be
\begin{equation}
 f_i = \biggl[1+ {3 \over 4} 
{{(1 + 2\, \zeta_\circ)(1 + \alpha (1 + 2\, \zeta_\circ))^2} \over
{\zeta_\circ + \sqrt{\zeta_\circ^2 + 3\zeta_\circ/8}}}\biggr]^{-1},
\end{equation}
where  $\alpha \equiv {\dot M}_w/{\dot M}_i$ 
is the ratio of wind to infall mass transfer rates.
This result is taken from equation (I.60), where the three terms of that
equation correspond, from left to right,  
to the normal force due to wind, infall, and gravity, respectively. 
For a fixed value of $\zeta_\circ$, decreasing $\alpha$ increases $f_i$. 
Thus, we consider the limit $\alpha \rightarrow 0$, to obtain the largest 
possible value of $f_i$. 
If we set $\alpha=0$ in equation (45), then $f_i$ monotonically 
increases with $\zeta_\circ$. The maximum of
$f_i$ occurs in the limit $\zeta_\circ \rightarrow \infty$, 
$\alpha\rightarrow 0$, in which case $f_i=4/7$. 
Since this is substantially 
less than the minimum value of $f_i$ necessary for stability, 
we conclude that there are no combinations of $\alpha$ and $\zeta_\circ$ 
such that the shell is stable. While this analysis is based upon
conditions near the symmetry axis, the small contribution of centrifugal
effects at larger angles (Section I.3.3)  suggests that the same result 
should hold throughout the shell.  
One might argue that the steady-state solution might nevertheless be 
achieved in some average sense. While this has been found to be true for  
isothermal bowshocks (Blondin \& Koerwer 1998; Raga \etal 1997), the 
calculations we next describe indicate that this is not the case in our 
problem.

\subsection{Numerical Demonstration}


We next use our time-dependent code to evolve shells starting near 
the steady-state 
conditions derived in Paper I. 
The latter calculation gives us the run of 
$R,\sigma,u_t,u_\phi$ for a grid of $\theta-$values. 
We first verified the 
validity of the steady-state solutions, checking that  the net
normal force on a fluid element is equal to the normal component of the 
centrifugal force it experiences. 
Despite this balance,  the shells do not stay in their initial configuration, 
but evolve quickly. 
Shells begun  at a size  slightly larger than  equilibrium expand,  
while those of slightly smaller size  
collapse, as expected by the analytical argument. 
A representative example is shown in 
Figure~\ref{fig:instab}. 
The innermost
two curves are polar radii for shells distorted slightly inward from
equilibrium, while the outermost two curves represent shells begun 
slightly beyond the equilibrium radius. 
All other quantities in the shell, such as the run of surface density
with polar angle, 
were initially the same as for steady-state. 
The equilibrium solution shown
is that of the critical (innermost) model for $\alpha=0.1$ shown in 
Figure I.9. 
The slightly shrunken, collapsing shell quickly accelerates to free-fall 
speed, 
while the expanding structure 
eventually decelerates at large radius
as a momentum-conserving snowplow. 
The analytic argument   
of the preceeding section, together with these and similar numerical 
experiments, 
show convincingly that the quasi-steady shells are 
dynamically unstable. This basic result 
motivates our current study, where we consider the shell evolution as an
explicitly time-dependent, initial value problem.

\section {Time-Dependent Evolution: Basic Considerations}
\subsection{Initial Conditions}

We now wish to follow the motion of the shell from its launch near 
the protostellar surface. 
Our shell is initially spherical and massless, with radius equal to that
of the protostar, $R_*$. 
As the structure evolves, it receives mass from both the
wind and infall. Thus, even at launch, 
there are well-defined values for all the physical quantities of interest. 
Note, however, that the momentum equation (37) is singular if $\sigma$ 
vanishes. The  initial velocity is found by setting the 
corresponding numerator to zero:
\begin{equation}
\rho_w\, {u}_{w,n}'\, ({\bf u}_w - {\bf u}_\circ) 
 - \rho_i\, {u}_{i,n}'\, ({\bf u}_i-{\bf u}_\circ)=0.\end{equation}
Here we have explicitly written the vector velocity differences. 
We define  
$\lambda \equiv - \rho_i u_{i,n}'/\rho_w u_{w,n}'$,
which is the ratio of mass fluxes 
from the wind and infall onto the shell.
Then equation (46) becomes 
\begin{equation}
 {\bf u}_\circ = {{ {\bf u}_w + \lambda\,{\bf u}_i}  \over
{1\, +\, \lambda}}.\end{equation}
To find $\lambda$ explicitly, we take the  normal  component of 
 equation (46), obtaining 
$\rho_w u_{w,n}'^2 = \rho_i u_{i,n}'^2$. 
Defining $\eta \equiv \rho_i/\rho_w$ 
and taking the square root of both sides, we have  
\begin{equation} (u_{w,n} - u_{\circ,n}) = 
\eta^{1/2} (u_{\circ,n}-u_{i,n}).\end{equation}
In obtaining equation (48), we have accounted for the fact that 
$u_{i,n}<u_{\circ,n}<u_{w,n}$. Equation (48) implies 
\begin{equation}\lambda =  \eta^{1/2},\end{equation}
which yields the final result
\begin{equation}
 {\bf u}_\circ = {{ {\bf u}_w + \eta^{1/2}\,{\bf u}_i}  \over
{1\, +\, \eta^{1/2}}}.\end{equation}
In equation (50), both ${\bf u}_i$ and $\eta$  depend on $\theta$. 
The normal component of this equation represents the balance of wind and
infall ram pressures in the frame of the shell. 
This result is a generalization of the 
well-known formula for the speed of the planar bow shock 
driven by a steady jet (e.g., Blandford, Begelman \& Rees 1984). 
In our case, however, the ``ambient'' matter is in nonuniform 
motion with velocity ${\bf u}_i$. 
For a dense wind $\eta \ll 1$, the shell
tends to the wind speed, while for a rarefied wind with $\eta \gg 1$, 
the shell collapses at the infall speed. 

To derive the initial rate at which mass is being swept up, 
we use equation (30):
\begin{eqnarray}
 {{D\,\sigma}\over {D\,t}}  \approx 
{{\partial\,\sigma} \over {\partial\,t}} & \equiv &
{\dot \sigma}_0,\nonumber\\
& = &
\rho_w\,u_{w,n}'-\rho_i\,u_{i,n}',\end{eqnarray}
where we have dropped terms from equation (30)
that  vanish with $\sigma$. 
The righthand side of equation (51) is to be evaluated 
using equation (50) for the shell's initial velocity.

Our calculation requires values for 
the stellar radius $R_*$, the wind speed 
$V_w \equiv |{\bf u}_w|$, and mass loss rate ${\dot M}_w$. 
The distribution of infalling matter is specified by  $t$,
the time since the start of collapse,
together with the sound speed $a_\circ$ and rotation rate $\Omega_\circ$ 
of the
parent dense core. 
Thus, there are six dimensional parameters.  
In principle, the protostellar radius is not independent of the others, but 
should be obtained from stellar evolution theory (e.g., Stahler 1988). 
However, we will simply adopt a 
characteristic protostellar radius and show how the results scale with
this and with other choices of the dimensional parameters.

\subsection{Nondimensional Parameters and Equations}

To reduce the number of runs needed to explore parameter
space, we cast the equations in nondimensional form.
There are two time scales of interest -- an {\it evolutionary} and a {\it
dynamical} one. The first is the time over which the infall itself
changes appreciably. In terms of the dimensional parameters listed
previously, we define
\begin{equation}
 t_{ev} \equiv \biggl({{R_*}\over {a_\circ\,\Omega_\circ^2}}\biggr)^{1/3}.
\end{equation}
According to equation (10) of Stahler \etal (1994), $t_{ev}$ is, to within a
factor of order unity, the time after start of collapse when the
centrifugal radius associated with infall attains the value $R_\ast$.

We find the dynamical time by first assigning units of length and
velocity. The first is $R_\ast$, and the second is the Keplerian speed at
the stellar surface. Since the characteristic stellar mass is ${\dot M}_i
\,t_{ev}$, we have 
\begin{equation}
V_{*}  \equiv  {{a_\circ^{4/3}} \over{R_*^{1/3}\,\Omega_\circ^{1/3}}},
\end{equation}
where we have left out the factor of $m_\circ$ entering equation (1) for 
${\dot M}_i$. The dynamical time is now defined as $R_*/V_*$, or 
\begin{equation}
t_{dyn} \equiv 
{{R_*^{4/3}\,\Omega_\circ^{1/3}}\over {a_\circ^{4/3}}}.\end{equation}
Note finally that the mass of the shell, after a dynamical time, 
is of magnitude  ${\dot M}_i\,t_{dyn}$. We formally define our mass unit as 
\begin{equation}
M_{dyn} \equiv {{a_\circ^3}\over {4\,\pi\,G\,\Omega_\circ}} 
\biggl({{R_*\,\Omega_\circ}\over {a_\circ}}\biggr)^{4/3}.
\end{equation}
Here again we have left out the factor of $m_\circ$, and have introduced
a factor of $4\,\pi$ for later convenience.

All physical variables can be written as dimensionless factors times
appropriate combinations of $R_\ast$, $t_{dyn}$, and $M_{dyn}$. Additionally,
certain  nondimensional ratios enter the final equations.
One involves the mass transport rates in the wind and infall:
\begin{equation} 
\alpha \equiv {{\Mdot_w}\over{\Mdot_i}} \,\,.
\end{equation} 
In this study, we only consider values of $\alpha$ less than unity. A
second ratio is the wind speed divided by our dynamical unit of
velocity:
\begin{equation} 
   \nu \equiv  {{V_w} \over {V_{*}}}.\end{equation}
Assigning  canonical values of the dimensional 
parameters $a_\circ=0.2\,{\rm km}\,{\rm s}^{-1}$,
$\Omega_\circ=2\times 10^{-14}\,{\rm s}^{-1}$, and $R_*=3\,R_\odot$, 
and a reference wind speed of $200\,{\rm km}\,{\rm s}^{-1}$, 
the reader may verify that 
$\nu$ is of order unity.
The third key nondimensional ratio is the initial launch time of
the shell in terms of $t_{ev}$:
\begin{equation}\tau_\circ \equiv  t_{init}/t_{ev}.\end{equation}
For $\tau_\circ\ll 1$, the centrifugal radius has not yet reached the 
protostellar radius, and the infall conditions are those of 
nearly spherically-symmetric free-fall. 
For $\tau_\circ \gg 1$, the initial launch is within
the inner,  anisotropic region of the infall. In order to compute
the evolution of the shocked shell in dimensionless units, only the
three nondimensional ratios $\alpha$, $\nu$, and $\tau_\circ$ need to 
be specified. 

Another  obvious ratio is that of the dynamical and evolutionary timescales:
\begin{eqnarray}   \epsilon  &\equiv&  {{t_{dyn}} \over {t_{ev}}},\nonumber\\
           &       = &    {{R_\ast\,\Omega_0} \over {a_o}}.\end{eqnarray}
This quantity is of order $10^{-7}$. 
Since our calculations span only
dynamical times, $\epsilon$ does not appear explicitly in the final,
nondimensional equations (see below). 
Note that our previous physical units can be written as
\begin{mathletters}
\begin{eqnarray}
    V_{*}  &=&  \epsilon^{-1/3}\, a_\circ,\\
     t_{dyn} &=&  \epsilon^{4/3}\, \Omega_\circ^{-1},\\
     M_{dyn} &=&\epsilon^{4/3}\, {{a_\circ^3}\over {4\,\pi\,G\,\Omega_\circ}}.\end{eqnarray}
\end{mathletters}

The nondimensional equations include various quantities related to the
infall. These, in turn, depend on the time $t$ since the start of
collapse.  Since our basic temporal unit is $t_{dyn}$, we may write 
\begin{equation}
t = \epsilon^{-1}\,\tau\,t_{dyn}.
\end{equation}
Here, $\tau$ is the nondimensional evolutionary time, written in 
terms of $t_{ev}$. In practice, the difference between $\tau$ and 
$\tau_\circ$ is exceedingly small. 
Even if we were to evolve the shell  for 
$10^4$ dynamical times, 
$\tau$ would only change fractionally  by order $10^{-3}$. 
Thus, in the dynamical equations we treat $\tau$ as a constant, and drop
the unnecessary subscript on $\tau_\circ$. 

To complete our description of the nondimensional problem, we give the  
dimensionless forms of the remaining infall and wind quantities. 
We write  the infall vector velocity in
terms of the Keplerian speed as ${\bf u}_i = u_K \,{\bf f}$, where
$u_K\equiv  {\sqrt{G\,M_*/R}}$, and the components of 
${\bf f}$ may be deduced from equations (3)-(5). 
Letting tildes denote nondimensional quantities 
(e.g. ${\tilde R}\equiv R/R_*$), and using $G\,M_* = m_\circ\, 
a_\circ^3\,t$, we obtain 
\begin{equation}  {\tilde u}_K = m_\circ^{1/2}\,\biggl({{\tau}
\over {\tilde R}}\biggr)^{1/2}.\end{equation}
Similarly, the quantity $\zeta$ is given by 
\begin{equation} \zeta = 
m_\circ^{3}\,{{\tau^3}
\over {16\,{\tilde R}}}.\end{equation}
When the centrifugal radius equals the stellar radius, $\zeta = 1$ at the
surface of the star. From the above equation, the corresponding nondimensional
time $\tau$ is $2.54$.
The infall and wind densities in nondimensional form are
\begin{eqnarray}  {\tilde \rho}_i & = &
{{{\tilde S}_i}\over {{\tilde R}^2\, {\tilde u}_K}},\\
{\tilde \rho}_w  & = &  {{\alpha}\over {{\tilde R}^2\, \nu}},
\end{eqnarray}
where the infall density shape function is  
\begin{equation}
{\tilde S}_i^{-1} \equiv 
- f_r\,\left[1+2\,\zeta\,P_2\left({\rm cos}\,\theta_\circ\right)\right].
\end{equation}

Dropping the tilde notation, the 
dimensionless form of the dynamical equations is 
\begin{eqnarray}
{D{\bf R}} \over {D\,{t}} &  = & {\bf u},\\
{{D\,({\rm ln}({\cal A}\,\sigma))} \over {D\,{t}}}
 & = &  {{1}\over 
{R^2\,\sigma}}
\biggl[\alpha \biggl(\cos\gamma-{{u_n}\over {\nu}}\biggr)-S_i\,\biggl(f_n-{{u_n}\over {u_K}}\biggr)
\biggr]\nonumber\\
  - {{\partial} \over {\partial \theta}}\biggl({{u_\theta} \over
{R}}\biggr),\\
{{D\,{\bf u}} \over {D\,{t}}} & =  & 
{{1}\over {R^2\,\sigma}}
\biggl[\alpha \biggl(\cos\gamma-{{u_n}\over {\nu}}\biggr)  {\bf u}_w'
-S_i\,\biggl(f_n-{{u_n}\over {u_K}}\biggr) {\bf u}_i'\biggr]\nonumber\\
 - m_\circ\,{{\tau\,
{\hat {\bf r}}}\over {{R}^2}}.\end{eqnarray}
Note again that only three parameters, the dimensionless ratios
 $\alpha$, $\tau$, and $\nu$, enter the final equations. 

In order to launch a shell, the driving wind speed must be sufficient
to yield a shell velocity $u_r>0$, which in turn implies 
that the wind ram pressure exceed the infall ram pressure at the stellar 
surface. In terms of our chosen units, we require $\nu>\nu_{ram}$, where
\begin{equation}
\nu_{ram}={{\nu_{esc}}\over {\alpha\,(1+m_\circ^3\,\tau^3/8)}}
\end{equation}
is the value of $\nu$ that balances ram pressures at the stellar pole, 
and  $\nu_{esc} = \sqrt{2\,m_\circ\,\tau}$ (c.f.~eq.[62]) is the escape
speed at the stellar surface.

\section{Numerical Results}
\subsection{Breakout versus Recollapse}

Before discussing the  details of our numerical results, we note that the 
mass of the shell typically will be dominated by the swept-up infall.
In the limit of a stationary, spherical shell, the ratio of wind to infall 
contributions is $\alpha \ (<1)$, but outward motion of the shell further  
increases the fractional contribution of infall to the shell mass. 
Although our calculations conserve the z-angular momentum of the incident,
infalling gas, the centrifugal support of the shell is modest. The
primary effect of infall angular momentum is to establish an asymmetric
flow,  which is then swept up by the expanding shell. 

A sample evolutionary sequence is shown in  
Figures~\ref{fig:breaking}-\ref{fig:breaking3} for a typical 
choice of nondimensional parameters ($\alpha=1/3$, 
$\tau=4$, $\nu=4.7$). 
Here, the results are displayed both nondimensionally and in physical units.
For the latter, we assumed standard values of the dimensional quantities:
$R_*=3\,R_\odot$,\ $\Omega_\circ=2\,\times\,10^{-14}\,{\rm s}^{-1}$,
$a_\circ=0.2\ {\rm km}\ {\rm s}^{-1}$. 
These choices imply an accretion rate of $1.85\times 10^{-6}\,{\rm M}_\odot\, {\rm yr}^{-1}$, a time since protostar formation of
$38,000$ years, and a wind speed of $159\  {\rm km}\ {\rm s}^{-1}$.
The shell first elongates as it expands up to and beyond the disk radius, 
indicated in  Figure~\ref{fig:breaking}. 
This outward motion never ceases, as the wind is able to drive back 
the infalling envelope in all directions. 
We refer to such an  outcome as {\it breakout}. 
While at intermediate times (see Figure~\ref{fig:breaking2})
the shell is quite elongated and a bipolar geometry is obtained, 
at late times  (Figure~\ref{fig:breaking3}) the shell takes 
on a more spherical appearance.
In this spatial regime, well beyond the disk radius, 
the ambient density is also spherically symmetric.

Figure~\ref{fig:rising} 
shows that dramatically different results are obtained with
very similar input parameters ($\alpha = 1/3$, $\tau = 4$, $\nu = 4.4$).
Dimensionally, the wind speed has now been lowered to  
$148\  {\rm km}\ {\rm s}^{-1}$. The shell initially
rises as before, but it stalls and falls back to 
the star at nearly free-fall speed. 
As the shell plummets to the stellar surface, initial
ripples are amplified; higher numerical resolution would be needed to follow
this apparent instability. We will not be interested in such 
details, but only in whether or not the shell succeeds in driving back
the infall. Shells that break out are typically quite smooth and
regular. Even for those that recollapse, the structure is smooth
up to the point of turnaround.  
Note that recollapse is caused primarily by 
gravity acting on the heavy shell, 
rather than infall ram pressure. Turning off gravity in the code, breakout is
achieved for a wind speed less than half of that in Figure~\ref{fig:rising}.

For those shells that do achieve breakout
and have traveled far relative to the initial stellar radius, all quantities
follow a power-law determined by the slope
in the infall density law. For $R_{cen}\,\ll\, r\,\ll\, a_\circ t$, 
$\rho_i\propto r^{-3/2}$, so the swept-up mass varies as $r^{3/2}$,
while the shell's momentum grows linearly with time. The resulting
momentum-driven snowplow has $R\propto t^{4/5}$ and $V_r\propto t^{-1/5}$.

\subsection{Critical Solutions}

We have mapped the parameter space 
of solutions as a function of $\alpha$, $\tau$, and $\nu$. 
For a given value of $\alpha$, 
the $\tau-\nu$ plane is divided into three regions:
one in which shells cannot even advance beyond the stellar surface, one in
which the shell may initially rise, but is pulled back, and one in which
the shell breaks out. These solution regimes will be called the crushed 
wind, the trapped wind, and the escaping wind, respectively. 
In the $\tau-\nu$ plane,
the boundary between the crushed wind and the trapped wind is the locus 
$\nu_{ram}(\alpha,\tau)$ given by equation (70), 
where infall and wind ram pressures balance at the stellar surface. 
We similarly  denote by $\nu_{crit}$ the minimum nondimensional 
wind speed necessary for breakout. 
Figure~\ref{fig:crits}
plots this quantity as a function of time, for three representative values of
$\alpha$. Again, we display the results both nondimensionally and in physical
units, employing our fiducial input parameters. Table~1 also lists values of
$\nu_{crit}$ at selected times. 
Returning to the figure, we see that, at very early times, 
$\nu_{crit}$ increases as $\tau^{1/2}$, and is also {\it inversely}
proportional to $\alpha$. 
At intermediate times such that $R_{cen}\,\sim\,R_*$, 
the centrifugal deflection of infalling gas lowers the infall 
density, and $\nu_{crit}$ consequently falls. 
Once again, $\nu_{crit}$ at any time during this epoch is proportional
to $\alpha^{-1}$. 
Throughout early and intermediate times, 
whether or not the shell breaks out is 
determined by the product $\alpha\, \nu$,
that is, on the momentum loss rate of the wind.
Finally, $\nu_{crit}$ again rises at late times. 
In this case,  the critical wind speed is  independent 
of $\alpha$ and increases as $\tau^{1/2}$.

Let us see in more detail how these results arise. 
The wind must combat the infall ram pressure and the gravitational 
force on the shell.  At early times, we may neglect  centrifugal distortion 
of the infall and consider only the  spherically symmetric problem. 
In this case, the infall ram pressure at fixed radius varies as $\tau^{1/2}$.
(Recall equation (39).)  
The gravitational force per unit shell mass rises as $\tau^1$. 
To determine the force per area, 
we first note that the infall density $\rho_i$ falls as $\tau^{-1/2}$, 
a consequence of the rising freefall velocity. 
For  a light wind, $\rho_w\ll\rho_i$,  
we may neglect the mass contribution of the wind. 
Since most of the mass of the shell is swept-up,
$\sigma\propto\rho_i\propto\,\tau^{-1/2}$, (c.f., equations (62), (64))
so the gravitational force per unit area on the shell  $\sigma\,{\cal V}$ 
also behaves as $\tau^{1/2}$. 
Comparing the wind ram pressure $\propto \alpha \,\nu$ to the sum of 
infall ram pressure and gravity, 
we obtain $\nu_{crit}\,\propto\,\tau^{1/2}/\alpha$.

The downturn in $\nu_{crit}$ 
at $\tau\sim 1$ occurs because the star is now interior to the centrifugal 
radius.   The decrease in the infall density along the z-axis relative 
to spherical accretion implies that breakout becomes easier. 
Because the shell's mass is dominated by the infall contribution, the decrease
in the infall density along the axis decreases both the infall ram pressure 
and the gravitational force on the shell. 
At late times, the centrifugal radius has grown so large that the primary 
source of mass input to the shell is the wind. 
In this regime, the wind speed necessary for breakout is proportional 
to the stellar escape velocity, which in turn varies as $\tau^{1/2}$.

We compare these numerical results for $\nu_{crit}$ with three 
analytical approximations in Figure~\ref{fig:analytic}. A 
hypothetical wind of speed $\nu=\nu_{esc}$ 
(the diagonal, dash-dot line) would be fast enough to break out 
if we neglect the mass and 
momentum fluxes to the shell from infall. We see that except
for late times ($\tau \ge 10$), the wind must be considerably 
faster than the escape speed to drive a swept-up shell that will 
break out. At late times, our constant speed wind can break out 
with $\nu<\nu_{esc}$ because we have neglected gravitational 
deceleration of the preshock wind.\footnote{Using modified wind conditions 
of such a coasting, decelerating wind of constant specific 
energy $e = V_w^2/2-G\,M_*/r$, and numerically determining 
the corresponding critical speed for breakout with our code 
(filled triangles  in Figure~\ref{fig:analytic}),
we see that the late-time critical curves are shifted vertically and
converge to the escape speed.  At early times, since breakout speeds 
were already much larger than the escape speed, the difference 
between a constant speed and a  decelerating wind is negligible, 
and the critical curves are unchanged. Because no observed winds 
actually decelerate,  we consider the constant velocity curves 
to be more applicable.} 
The second approximation is
that of $\nu=\nu_{ram}$, shown by the dashed line, yielding ram
pressure balance of wind and infall at the stellar surface. 
This approximation underestimates the necessary wind speed for breakout 
by a factor of $1.9$ at early times, when the infall is spherically symmetric,
but it underestimates the critical speed by a much larger factor
once the infall becomes significantly aspherical. 

The two foregoing analytical approximations give a poor agreement 
with $\nu_{crit}$ because the first
neglects the infall ram pressure, while the second neglects the gravitational
force on the shell. To include in an analytic fashion the effect 
of both infall ram  pressure and gravity on the shocked gas within the shell, 
we specify the condition that the initial
{\it postshock} gas be rising at the escape speed. 
According to equation~(50),
in order to have $u_{\circ,r}=V_{esc}$, we obtain the necessary condition
\begin{equation}\nu=\nu_{esc} (1+2\,\eta^{1/2}).\end{equation}
We cannot immediately apply this equation, because $\eta=\rho_i/\rho_w$ 
itself depends 
on the wind speed. The density 
ratio is found from equations (64)-(66), where for
$\theta=0$ we have $f_r = -\sqrt{2}$ and $P_2(\cos\theta_\circ)=1$, 
which yields
\begin{equation}
\eta = b^2\,{{\nu} \over {\nu_{esc}}},
\end{equation} 
where we have defined for brevity 
\begin{equation}
b\,\equiv\,1/\sqrt{\alpha\,(1+2\,\zeta_\circ)}.
\end{equation}
The solution $\nu'$ to equations (71), (72) 
is found as the root of a quadratic equation, giving
\begin{equation}
\nu'= f\,\nu_{esc}\,[b+\sqrt{b^2+1}]^2,
\end{equation}
which for $f=1$  is the value of $\nu$ required to give the 
{\it postshock, mixed} 
gas escape speed at the stellar surface. The ad hoc factor 
$f$ allows us to consider the wind to be some fraction $f$ 
of the speed required to give the shell escape speed at launch. 
The corresponding curve for $\alpha=0.1$ and $f=1$ 
is shown in Figure~\ref{fig:analytic} (dotted curve). 
If we neglect changes in the infall momentum flux per unit solid angle
once the shell is launched, $\nu'$ 
is expected to be an estimate of the necessary
wind speed to launch a shell that will break out. In practice, as
the shell advances,  the forces on the shell change, and typically
the wind weakens less quickly than the infall, so the above condition
overestimates the required wind speed. Except for the vertical offset, 
however, the curve follows closely the shape of the numerically-derived 
critical curve, suggesting $f\approx constant$
throughout the evolution. By choosing $f=0.45$, so as to fit the late-time
behavior, an approximate fit to the critical curves is shown in 
Figure~\ref{fig:crits}.

\subsection{Generalization to Anisotropic Wind}

The previously described calculations may be extended to anisotropic 
winds by defining angular dependences of the wind mass and momentum 
fluxes according to
\begin{eqnarray}
\rho_w V_w = {{{\dot M}_w} \over {4 \pi\, r^2}}\,f_w(\theta),\\
 \rho_w V_w^2 = 
{{{\dot M}_w\,{\bar V}_w} \over {4 \pi \,r^2}} \,g_w(\theta),
\end{eqnarray}
where ${\bar V}_w$ is the streamline-averaged wind speed 
and the functions $f_w$ and $g_w$  are normalized to have 
unit average value  over $4\pi$ steradians. 
Although a diversity of shell shapes may be generated in this 
fashon, we focus only on the behavior at the symmetry axis,
where breakout is easiest. If the properties of the wind are smooth
near the pole, i.e. $f_w'(0)=g_w'(0)=0$, the shell has $R'(0)=0$
and looks locally as though it is driven by an isotropic wind. 

Figure~\ref{fig:breaking2} shows a numerical example. 
Here the dashed curves  represent a shell driven by an anisotropic wind 
specifically chosen to have the same mass and momentum fluxes towards 
$\theta=0$ as the isotropic wind (solid curves). 
We have chosen the asymmetric driving wind to have $\alpha=1/6$ 
and a density  $\rho_w \propto f_w = g_w = \frac{1}{2} (1+3\,\cos^2\theta)$, 
with $V_w$ independent of $\theta$, while 
the values of $\nu$ and $\tau$ are unchanged from the spherical wind case. 
This angular dependence represents the lowest-order expansion that is 
axisymmetric and  of even parity, and has been chosen arbitrarily to give  
a pole-to-equator density contrast of $4$. 
Because the derived polar behavior $R_\circ(t)$ of a shell driven by an 
anisotropic wind follows that of a  spherical wind having
the same polar mass and momentum fluxes, the critical wind speed $\nu_{crit}$
for breakout of an anisotropic wind may be determined from our numerical 
curves. Defining  $\nu\equiv{\bar V}_w/V_*$,  the equivalent
spherical wind corresponds to 
$\alpha_{sph}=\alpha\, f_w(0)$ and $\nu_{sph}=\nu\, g_w(0)/f_w(0)$.
A wind that is focused preferentially towards
the poles mimics a spherical wind 
corresponding to a  higher value of $\alpha$ 
and a reduced value of $\nu_{crit}$, 
breaking out at an earlier time (see below).  
In our example, the  $\alpha=1/6$ wind 
has $\alpha_{sph}=1/3$ and the shell propagates along the axis like a bubble
from  an isotropic wind of higher $\alpha$-value.

\subsection{The Breakout Time}

Given the curves for the critical wind speed for breakout (Figure~9), we 
now wish to estimate the time at which breakout occurs. 
We first note that a wind whose speed is independent of $\tau$ would
experience no trapped phase, since the curves for $\nu_{crit}$ decrease
as $\tau\rightarrow 0$. 
Although we are not offering a general theoretical account of protostellar 
winds, we note that a constant wind speed is unlikely on purely empirical
grounds. Stars of all masses and ages tend to have $V_w$-values roughly equal
to the appropriate surface escape speed (Lamers \& Cassinelli 1999). 
We therefore recast Figure~9
as a plot of $V_w/V_{esc}  = \nu/\nu_{esc}$ versus $\tau$ (Figure~11). For
comparison, we have also plotted $\nu_{ram}/\nu_{esc}$, 
to delimit the end of the 
crushed wind phase, using equation (70) for $\nu_{ram}$. 
Suppose, then, that $\nu/\nu_{esc}$ were strictly a constant, equal to unity. 
Taking the case $\alpha = 1/3$ in Figure~11, we see that the wind would
first be able
to advance from the stellar surface at $t \sim 25,000$ years, but breakout 
would be achieved at $t \sim 54,000$ years. For the $\alpha=1/10$ 
model, these times increase to $41,000$ and $98,000$ years, respectively. 
Our  numbers rely on a relatively modest value of 
$\Omega_\circ=2\,\times\,10^{-14}\,{\rm s}^{-1}$. 
Based upon equation (52), increasing $\Omega_\circ$ by a factor of two would
decrease the above breakout times by a factor of $0.63$.
Thus, the trapped outflow phase may be a significant  fraction ($\sim 60\%$)
of the time prior to breakout.


For the example we have shown,  when the shell breaks out, it does so
at essentially all angles. However, this is not the case for all runs and is 
not true in general.
Thus,  our breakout time is to be interpreted as the time at which the 
wind first breaks out along the $z-$axis. At larger angles, the wind may 
be unable to break out until a later evolutionary time, resulting 
in simultaneous outflow and infall. 
Usually, however, the wind escapes globally, 
as shown in Figures~5-7.
The shell aspect ratio then begins and ends as unity, either at the stellar
surface or far from the star. The maximum distortion, which is still modest, 
is attained at intermediate times, when the shell is a few AU from the
stellar surface. 


\section{Discussion}

\subsection{Behavior during the Trapped Wind Phase}

Our recollapsing shells generally become unstable during the trapped wind 
phase. Portions that are initially inside neighboring regions fall faster, 
increasing any perturbation of an initially smooth surface. 
While our calculations do find this strong dynamical effect, 
we cannot detect the nonlinear instability of
Vishniac (1994), which relies on a finite shell thickness.
Our results make it clear, in any case, that the true behavior during 
the trapped wind phase is exceedingly complex. After a portion of the 
shell collapses to the stellar and disk surfaces, a new shock is 
immediately driven outward by the steady wind.
Thus, there will simultaneously be contracting and expanding patches at
different angles.  Portions of the wind may escape temporarily between 
recollapsing fragments before falling back. 
While the details of this process cannot be followed by our computational
method, we believe  our calculated breakout times should still be reasonably 
accurate. In any event, this time cannot decrease by more than about a 
factor of two, since the infall crushes the wind entirely at an earlier epoch.

One observational signature of the trapped wind phase would be fluctuations
in luminosity due to the rising and recollapse of the shell. This luminosity
arises from both the shocking of the wind and ambient gas. 
When a portion of the shell is rising, 
the corresponding wind component of luminosity decreases because of the
lowered shock velocity. 
The infall component of the luminosity may either increase 
(because the shock velocity is raised), 
or decrease (because $|{\bf u}_i|=u_{esc}$ decreases at larger radius.) 
All luminosity  fluctuations should increase in both period and amplitude 
with evolutionary time, because as breakout is  approached, 
the shell rises higher before recollapse.  Further study is needed 
to determine the observed magnitude of these fluctuations, as well as  
the transport of radiation  through the infalling gas.
As indicated earlier, the wind may also break out along the axis, 
while portions of the shell at larger angles continue oscillatory behavior. 
In this case, one would observe a bipolar jet accompanied by quasi-periodic 
fluctuations in luminosity. 

\subsection{Comments on Collimation}

Figures~5-7 show that the lobe-like appearance of our shells is a transient
phenomenon created by anisotropy in the infalling gas. In particular, this
elongated morphology cannot be identified with the collimation seen in 
Herbig-Haro jets. Our assumed mixing of shocked wind and infall is, in fact,
an agent for {\it decollimation}. A fluid element of the wind, initially
moving radially outward, acquires a positive $\theta$-velocity. Thus, there is
a tendency for gas to move towards the disk plane, despite the (temporary)
elongation of the shell.

Our mixing assumption was made purely for computational simplicity. 
In reality,
there must be considerable shear between the two shocks. This shear may have
several important consequences. First, shocked wind may be refracted 
along the 
shell's inner surface toward the polar axis, resulting in a more jet-like
flow. Such a geometry would resemble the early outflow model of Cant\'o
(1980), although the latter pictured a shell in pressure balance with 
a static, external medium. A second effect is that shocked, 
ambient gas will move tangentially toward the {\it equator}. 
The overall result will be a shell of
lower mass that is less confined by stellar gravity. The duration of the
trapped wind phase will therefore be reduced, resulting in earlier breakout.
However, we note that even in the limit of no mixing, the wind must still be
crushed by infall at a sufficiently early epoch. 
In future calculations, we
hope to explore quantitatively the consequences of relaxing the mixing
assumption. 

In a series of numerical simluations carried out through the 1990s, A.~Frank
and colleagues found that the interaction of a spherical wind and infall
produces very strong collimation. However, these studies (beginning with
Frank \& Noriega-Crespo 1994) 
modified the equations for infall, thus obtaining
a much larger density anisotropy than adopted here. For example, Frank \&
Mellema (1996) used an equator-to-pole density contrast of $50-70$, while
Delamarter \etal (2000) allow this ratio to exceed $10^3$. For comparison, 
the density contrast in our outflow only exceeds 50 only 
for $0.96 \lsim R/R_{cen} \lsim 1.08$. 
Our escaping shell spends relatively little 
time within this region, sweeping up a small amount of mass. We conclude that
a spherical or modestly anisotropic wind undergoes little collimation through
interaction with a physically realistic infall, unless it is through the
shear effect described above.

Our finding of an early, trapped wind phase is a similarly robust result, 
although it was missed in previous simulations. Frank \& Mellema  adopt
such a large wind velocity ($V_w \ge 500$ \kms) that the infall ram pressure
is overwhelmed. Mellema \& Frank (1997) did find oscillating shells
that, under some conditions, collapsed. However, the driver of their 
oscillations was a variable wind, so that collapse always coincided with the
weakest outflow phase. More recently, Delamarter \etal (2000) utilized as an
infalling background the flattened, rotating cloud of 
Hartmann, Calvet, \& Boss (1996).
Although they claim that especially weak winds may be stifled entirely by
infall, their numerical results always show the wind escaping through at least
a narrow solid angle through the symmetry axis.

We maintain, based on the arguments presented above, that early wind crushing
and eventual breakout are inescapable features within a realistic account of
the wind-infall interaction. In contrast, a strong, untrapped wind that is
present {\it ab initio}, as in previous simulations, would have already 
modified substantially the background infall, rendering subsequent results 
of dubious validity. Our own calculations of breakout can provide 
useful starting conditions for future investigators who wish to follow 
the wind evolution well beyond this critical, early phase.

\subsection{Further Observational Considerations}

We have seen that, for canonical values of our parameters, the wind is trapped
for a period of roughly $50,000$ yr. 
Although the evolutionary status of the most
embedded sources is by no means secure, those designated as Class 0 (Andr\'e
\etal 1993) are generally considered to be the youngest. They are detectable
only at far-infrared and submillimeter wavelengths, and have spectral energy
distributions corresponding to dust temperatures of roughly 30 K. 
An additional argument for their extreme youth is their relatively 
low population. In a
submillimeter survey of $\rho$ Ophiuchi, Motte \etal (1998) found that Class I
sources outnumber those of Class 0 by a factor of 10. 
Since the former have
traditionally been assigned ages of order $10^5$ yr (Kenyon \etal 1990), 
Class 0 stars could be 
as young as $10^4$ yr, if the population is forming in steady-state fashion. 
Visser \etal (2002), in their recent SCUBA survey of Lynds dark clouds, have
challenged these figures. They find nearly equal numbers of Class 0 and I
sources, and conclude that the former also have ages of roughly $10^5$ yr. 

Regardless of the outcome of this controversy, both surveys, as well as other
studies, have found that essentially all stars with massive, dusty envelopes
produce winds, as evidenced by associated CO outflows. 
Where, then, is the  trapped phase? 
We see two possible answers to this question. The first, and less
likely, is that our choice of input parameters requires adjustment. 
However, the necessary changes would be severe. 
For fixed $\alpha$, $\nu$, and $\tau$, $t_{break}$
scales as  $R_*^{1/3}\,a_\circ^{-1/3}\,\Omega^{-2/3}$. 
It strains credibility to suppose, for example, that the cloud 
sonic speed $a_\circ$ is so high that $t_{break}$ 
is reduced by an order of magnitude. 
In addition to this change of scale, we may change nondimensional solutions,
by assuming a faster wind (greater $\nu$). As seen in Figure~\ref{fig:nonsteady}, the breakout time is relatively insensitive to $\nu$ until a value is
reached where the trapped wind disappears entirely (where the dashed curves
are horizontal). This option, however, would imply very high wind speeds
($V_w \sim 3\,V_{esc}$ for $\alpha = 1/3$, or $V_w \sim 10\,V_{esc}$ for
$\alpha=1/10$). 
As noted in \S 6.3, making the wind more anisotropic effectively 
raises $\alpha$  and also leads to earlier breakout. 
Winds launched via the magnetocentrifugal 
mechanism, however, typically have modest anisotropy until they propagate a 
considerable distance (see, \eg, Najita \& Shu 1994). 
A second, and more plausible, answer is 
that we are {\it already} witnessing at least partially trapped winds. 
Following the arguments in \S 7.1, it may be that simultaneous infall and 
outflow occur well before our idealized model indicates 
the onset of breakout.  It would be extremely interesting, 
in this regard, to monitor the temporal 
variability of outflows from Class 0 sources.  

\section{Conclusions}

We have presented a detailed numerical calculation of the interaction between
a spherical, protostellar wind and an anisotropic, infalling envelope. The
anisotropy in our model is derived from the rotation of the star's parent 
cloud core. Collapse is assumed to proceed in inside-out fashon from a 
singular, isothermal sphere, in the absence of magnetic forces. 
We have idealized the double-shock interaction region as a thin
shell that is well-mixed internally. After demonstrating that our previous,
quasi-steady solutions are dynamically unstable, we have followed the shell
dynamics in a fully time-dependent manner. 

At very early times, the wind is crushed by ram pressure associated with
infalling matter. Wind breakout is delayed by the gravitational force exerted
on the shocked material. The shell must not only be supported by wind ram
pressure, but must have sufficent kinetic energy to escape the star's 
potential well, even while gathering additional mass from the envelope. 
Breakout would
occur earlier if either the wind or infall were more anisotropic, or if we
were to include the internal shear between the shock surfaces. 

Bearing in mind the observations of jets and outflows from embedded stars,
we acknowledge two important limitations of our model. First, our assumption 
of complete, early trapping may be incorrect, because of shell fragmentation.
Second, the observed collimation of stellar jets over long distances is not
approached asymptotically in our calculation. Instead, our shells, while they
are temporarily elongated, eventually become spherical. 
Jet collimation may be a consequence of magnetic pinching in the wind itself,
or of a cloud background that is very different from the one we
assume here. An initially more anisotropic cloud core will, of course, 
yield an altered pattern of infall. Alternatively, a shallower falloff
in the background density will give rise to crossing shocks in the 
interaction region (see, e.g., ~Cant\'o, Raga, \& Binette 1989). These
shocks further aid in wind collimation. 





\acknowledgments
FPW acknowledges helpful discussion with S.Lizano, and is grateful to the 
NSF International Researchers Fellowship Program 
and CONACyT/M\'exico for financial support. Part of the calculations were
performed at the Observatoire de la C\^ote d'Azur as a Henri Poincar\'e
Fellow.  SWS was supported by NSF Grant AST 99-87266. 



\begin{deluxetable}{llll}  
\tablecolumns{4}  
\tablewidth{0pc}  
\tablecaption{Critical Wind Speed $\nu_{crit}$}  
\tablehead{  \colhead{}    &  \multicolumn{3}{c}{$\alpha$}  \\  
\cline{2-4}  \\  
\colhead{$\tau$} & \colhead{$1/3$}   & \colhead{$1/10$}    & \colhead{$1/30$}}
\startdata  
    $0.25$   &$4.19$ &$12.7$ &$37.1$\\
    $0.50$   &$5.90$ &$17.9$ &$52.0$\\
    $1.00$   &$7.67$ &$23.1$ &$67.0$\\
    $2.00$   &$7.34$ &$21.0$ &$59.8$\\
    $4.00$   &$4.48$ &$11.6$ &$32.2$\\
    $8.00$   &$2.81$ &$5.80$ &$16.1$\\
    $16.0$   &$2.89$ &$3.33$ &$8.06$\\
    $32.0$   &$3.71$ &$3.87$ &$4.18$\\
    $64.0$   &$5.10$ &$5.16$ &$5.29$\\
    $128.$   &$7.14$ &$7.18$ &$7.23$\\
    $256.$   &$10.1$ &$10.1$ &$10.1$\\
\enddata 
\label{tab:table1} 
\end{deluxetable}


\clearpage
\setcounter{figure}{0}
\figcaption{Element of the shell, with normal 
and tangential directions shown. 
The angle $\gamma$ is measured clockwise from the radial to the normal 
direction. The shell thickness is $\Delta n$, while the arc length at
constant azimuthal angle is $\Delta s$. The azimuthal width of the patch is
$\Delta w = R \sin\theta \Delta \phi$. The length of the chord slicing
the shell at constant $(\theta,\phi)$ is $\Delta r = \Delta n \,\sec\gamma$.
\label{fig:wsIf4}}

\figcaption{Approach of the shell to the steady-state bow shock solution. 
In the frame of the star, located at the origin, the ambient medium 
moves with velocity $-V_*\, {\hat {\bf z}}$. The initially spherical shell 
began at $3\, R_{sun}$, much smaller than the standoff radius 
$R_\circ=2.45\times 10^{17}{\rm cm}$. Solutions
are shown for elapsed time increasing by a factor of two from 
$\frac{1}{32}$ to $4$
times the crossing time $R_\circ/V_*$. Triangles denote gridpoints, while the
analytic, steady-state solution is given by the solid curve.\label{fig:wsIf3}}

\figcaption{Fractional errors for the bow shock test case after reaching
near-equilibrium. Because the normal velocity vanishes in steady state, the 
quantity $u_n/u_t$ refers to the normal velocity of the gridpoints divided 
by the analytic solution for the tangential velocity.\label{fig:errors}}

\figcaption{Instability of an Initial Steady-State Solution: Polar radius 
vs. time for shells starting close to the steady-state solution of Paper I.
Shells begun at slightly smaller radius collapse, while shells begin
at slightly larger radius expand indefinitely. The equilibrium radius is 
shown (boldface) for the solution corresponding to $\alpha=0.1$, 
$\zeta_\circ=2.45$.\label{fig:instab}}

\figcaption{Time-evolution of an escaping shell (early evolution),
corresponding to $\alpha=1/3$, $\nu=4.7$, and $\tau=4.0$. 
The protostellar age since core formation 
is  $3.8\times\,10^4$ years. 
The shapes correspond to equal time intervals of $0.016$ years.
The subsequent evolution of this shell is shown in the
next two figures on a larger scale. The scale of the centrifugal
radius is indicated by the disk.  Lengths are in cm on the left and 
bottom axes,  and in units of the stellar radius on the top 
and right axes.\label{fig:breaking}}

\figcaption{Time-evolution of an escaping shell (further evolution),
corresponding to the same parameters as Figure~\ref{fig:breaking} 
but for with increasing
times steps and on a larger scale. 
The innermost curve of this figure is the same as the outermost 
of that Figure. However, the elapsed time increases by a factor 
of two with each curve: 4,8,16,32, and 64 time units. The dashed
curves represent an ``equivalent'' model driven by an asymmetric wind 
(See \S 6.3)\label{fig:breaking2}}

\figcaption{Time-evolution of an escaping shell (late time),
corresponding to the same parameters as Figures~\ref{fig:breaking} and 
\ref{fig:breaking2}. 
Here the shell has gone well beyond the centrifugal radius
of the infall and is in the increasingly spherically symmetric
infall region. Hence the shell becomes more spherical as it 
sweeps up this material. The elapsed time doubles with each successive
curve. The innermost here corresponds to  128 time units of 
Figure~\ref{fig:breaking}, while the outermost is greater by a
factor of 64 (8192 time units), which corresponds to
$134$ years since launch.\label{fig:breaking3}}

\figcaption{Time-evolution of a recollapsing shell 
corresponding to $\alpha=1/3$, $\nu=4.4$, and $\tau=4.0$.
The critical wind speed for these parameters is $\nu_{crit}=4.48$.
Solid curves show the rising phase,
while dotted curves display the subsequent recollapse). 
Shown for equal time intervals of $1/4$ the time to reach
the highest point along the z-axis, which is the same interval as the
unit used in Fig.~\ref{fig:breaking}.\label{fig:rising}}

\figcaption{Minimum breakout wind speed versus evolutionary time.
The three loci (solid curves) correspond to three ratios, $\alpha$, of the 
wind mass loss to infall accretion rate. For a given $\alpha$, 
the region above the curve corresponds to breakout, while that below
the curve corresponds to recollapse. 
An analytic fit to the numerical solutions is shown in 
dashed curves.\label{fig:crits}}

\figcaption{Minimum breakout wind speed versus evolutionary time, compared
with analytical arguments. All curves correspond to $\alpha=0.1$. 
At bottom left (dashed curve), the ram pressure balance
condition at the launch point. Bottom right (diagonal, dot-dashed line), 
the condition of initial wind speed equal to the escape speed. 
Curve labeled $\nu_{crit}$: 
numerical results for the critical wind speed. 
Top curve (dotted): the condition
that the shocked shell initially move at the escape speed. Filled triangles: 
critical condition for a decelerating wind (see text).\label{fig:analytic}}

\figcaption{Critical wind speed for breakout (solid curves),  in units 
of the free-fall (escape) speed, 
as a function of evolutionary time. The corresponding $\alpha$-values
are shown, as well as the wind speed necessary for ram pressure balance
at the stellar surface (dashed curves). Assuming wind launch conditions
$\nu_w/\nu_{esc}=1$ (i.e. following the horizontal line in this 
figure), evolution begins at the left edge of the plot with
the wind unable to advance beyond the stellar surface until 
the line intersects the appropriate dashed curve. Then 
the trapped wind phase lasts until the line intersects 
the corresponding solid curve for breakout.\label{fig:nonsteady}}

\newpage
\plotone{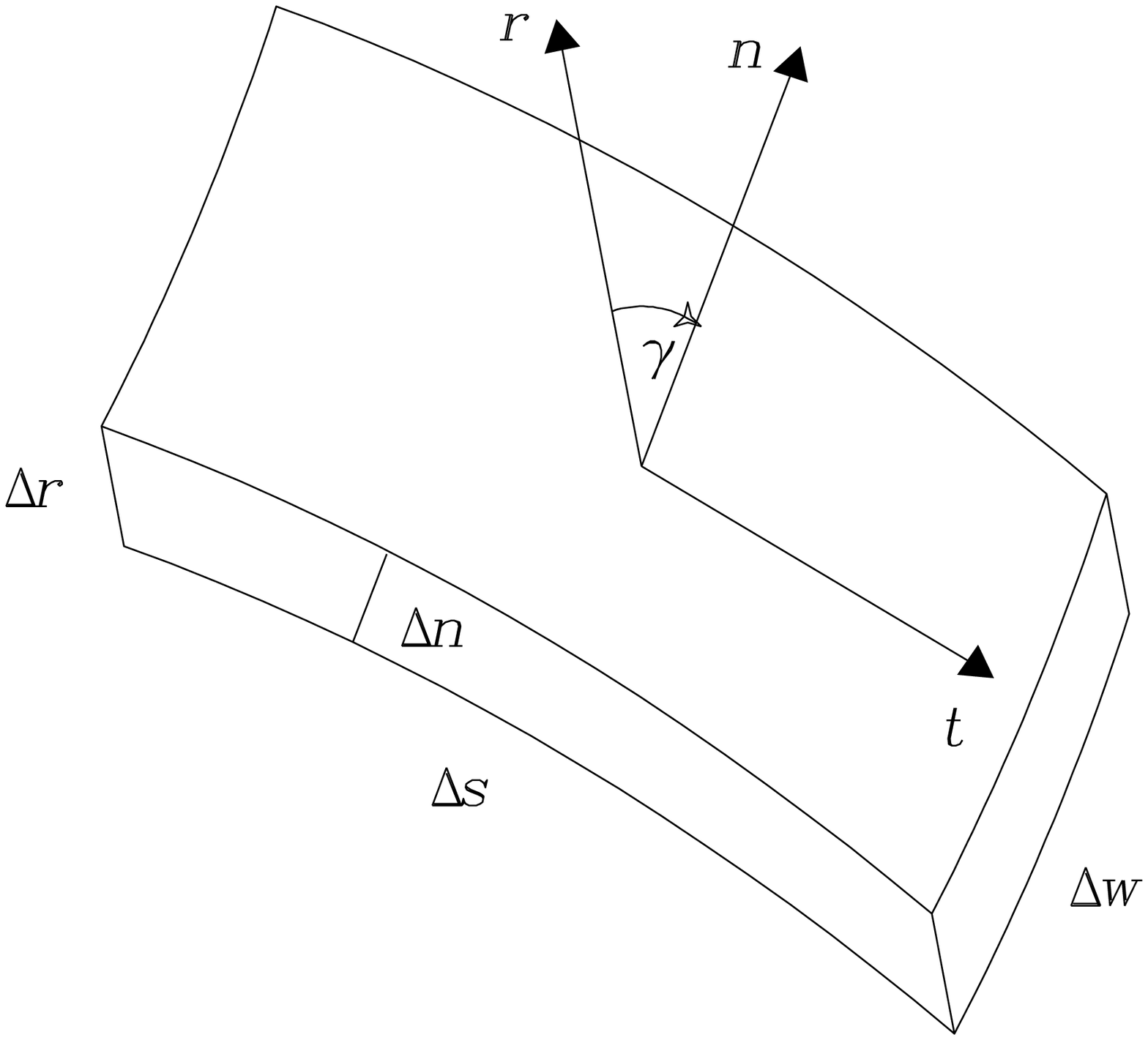}
\newpage
\plotone{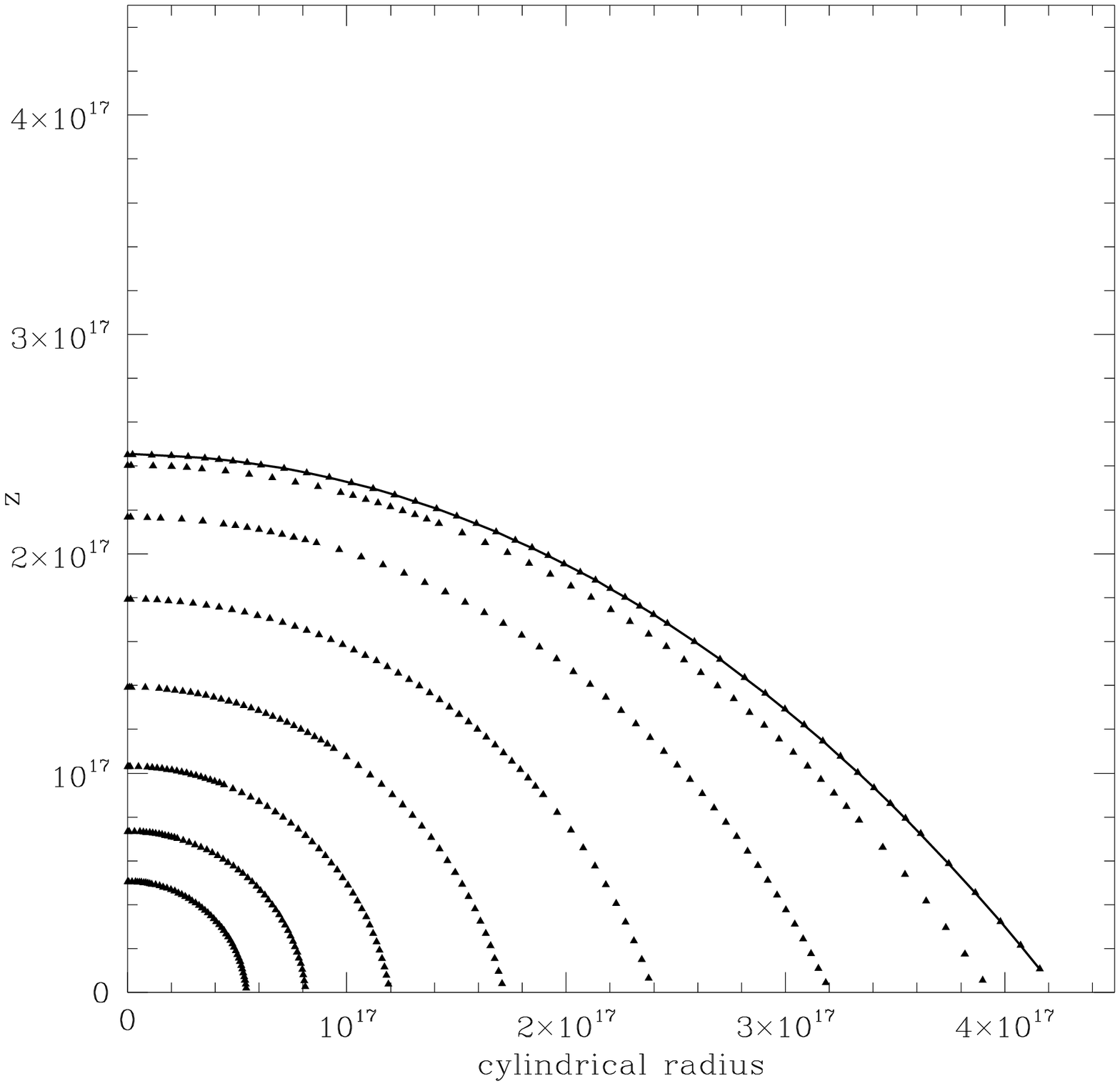}
\newpage
\plotone{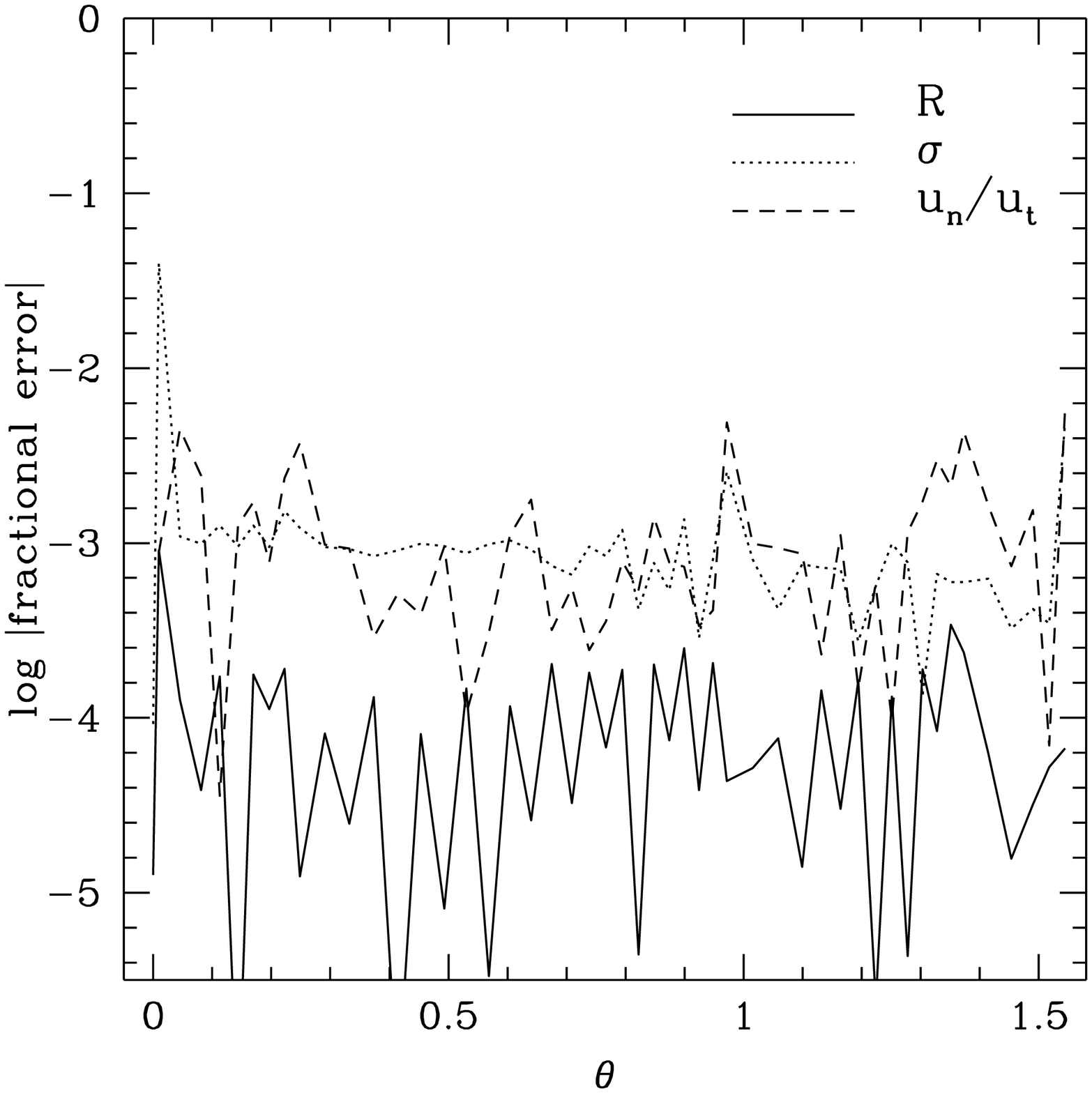}
\newpage
\plotone{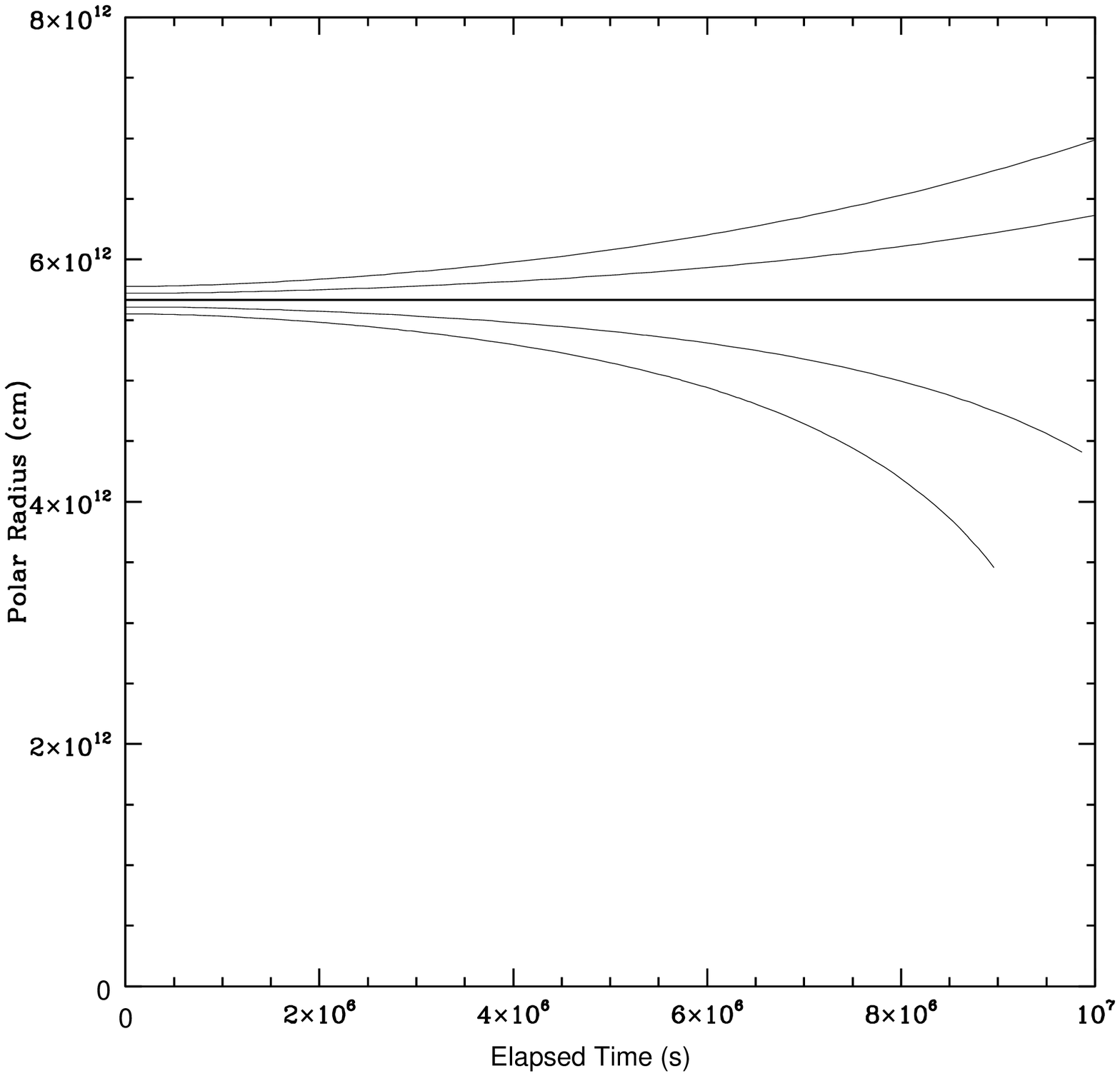}
\newpage
\plotone{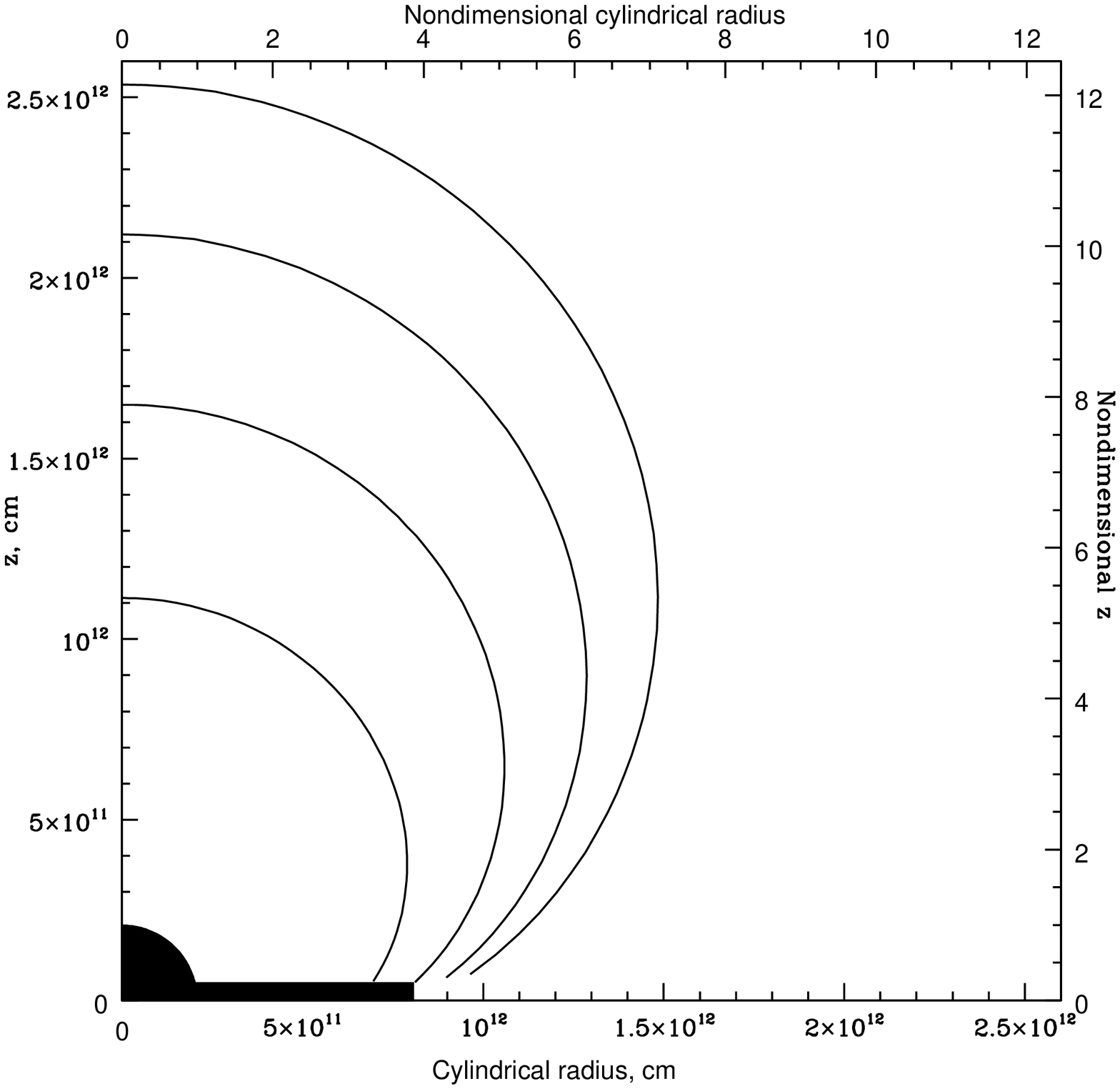}
\newpage
\plotone{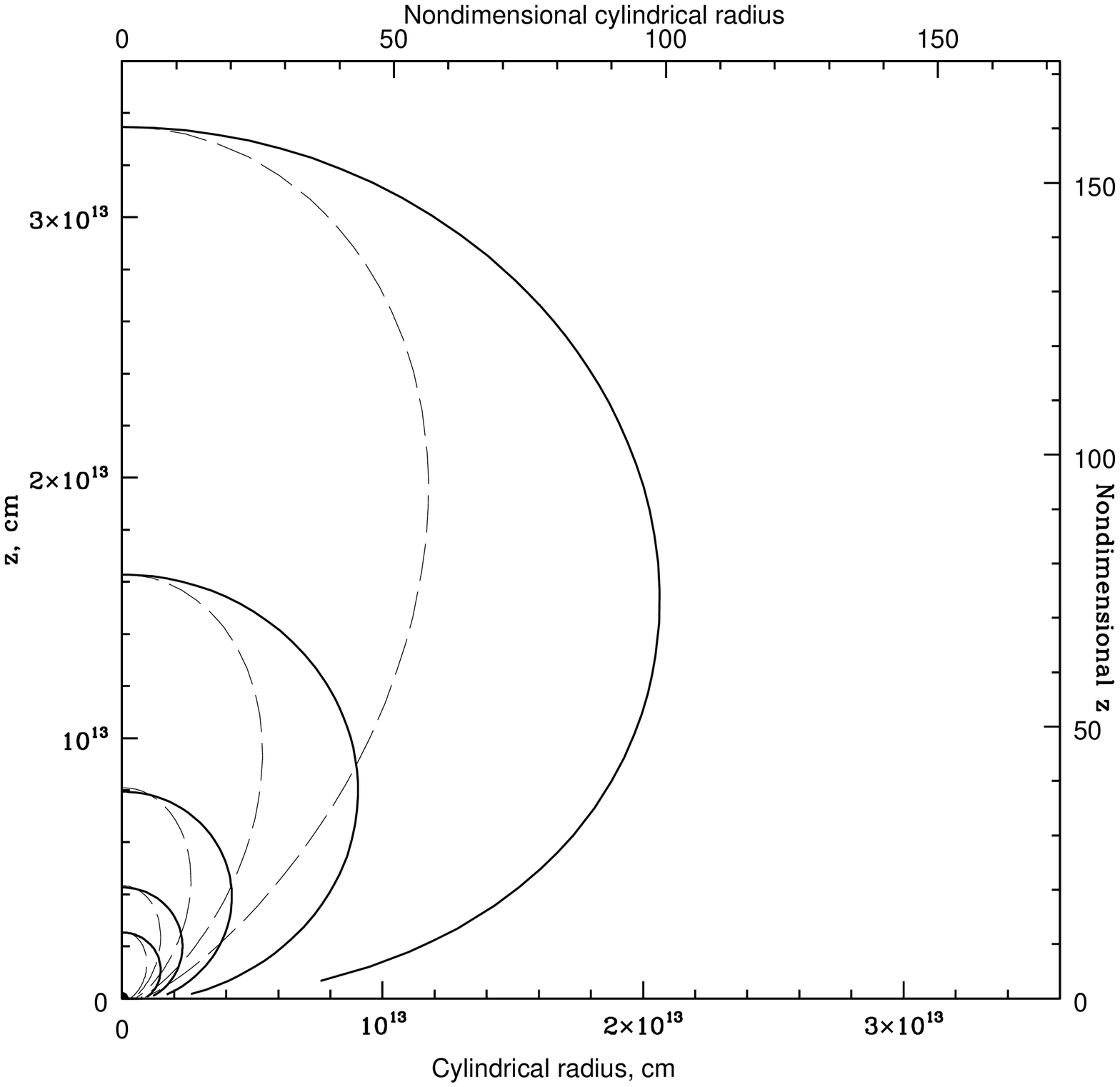}
\newpage
\plotone{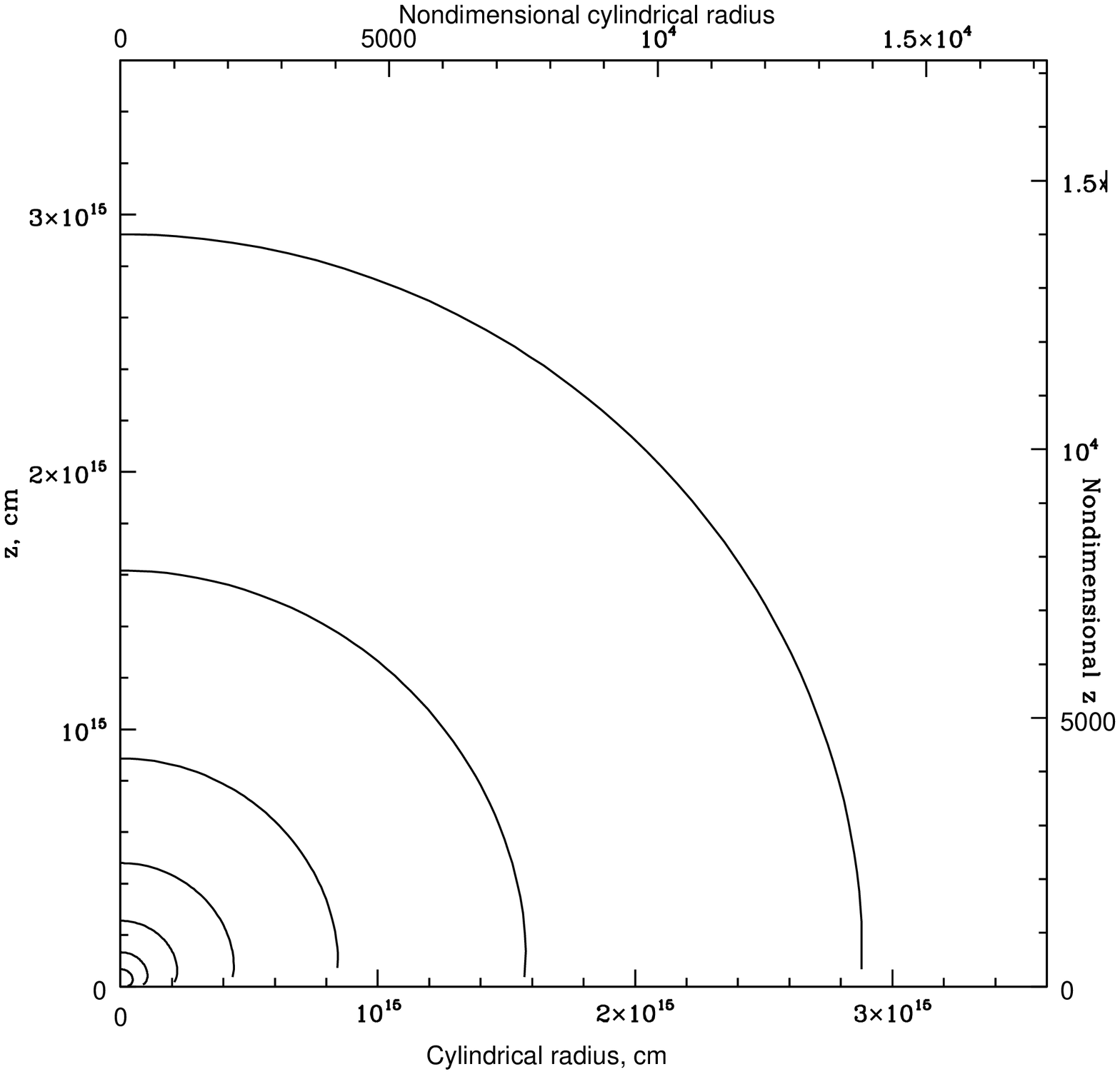}
\newpage
\plotone{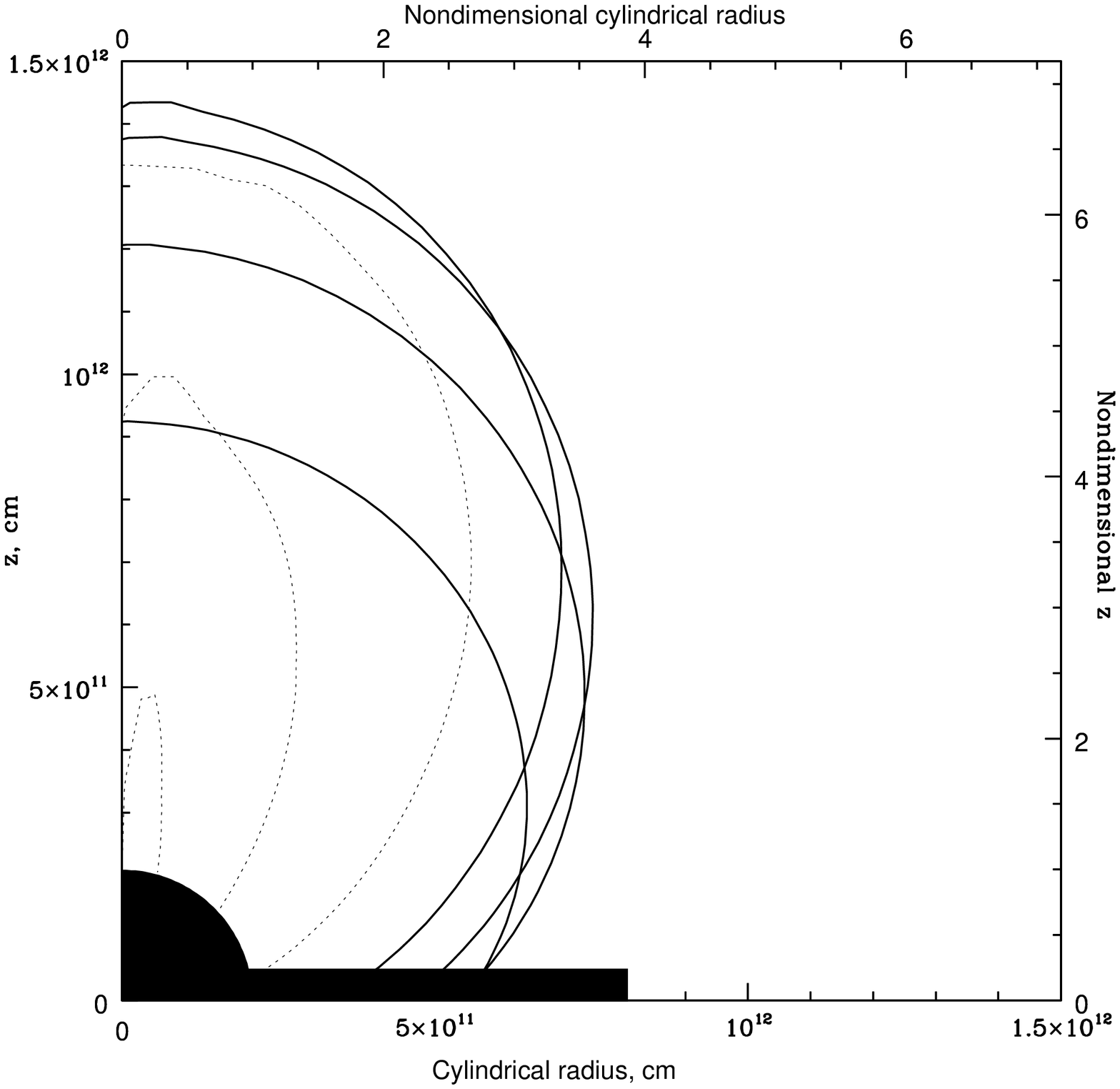}
\newpage
\plotone{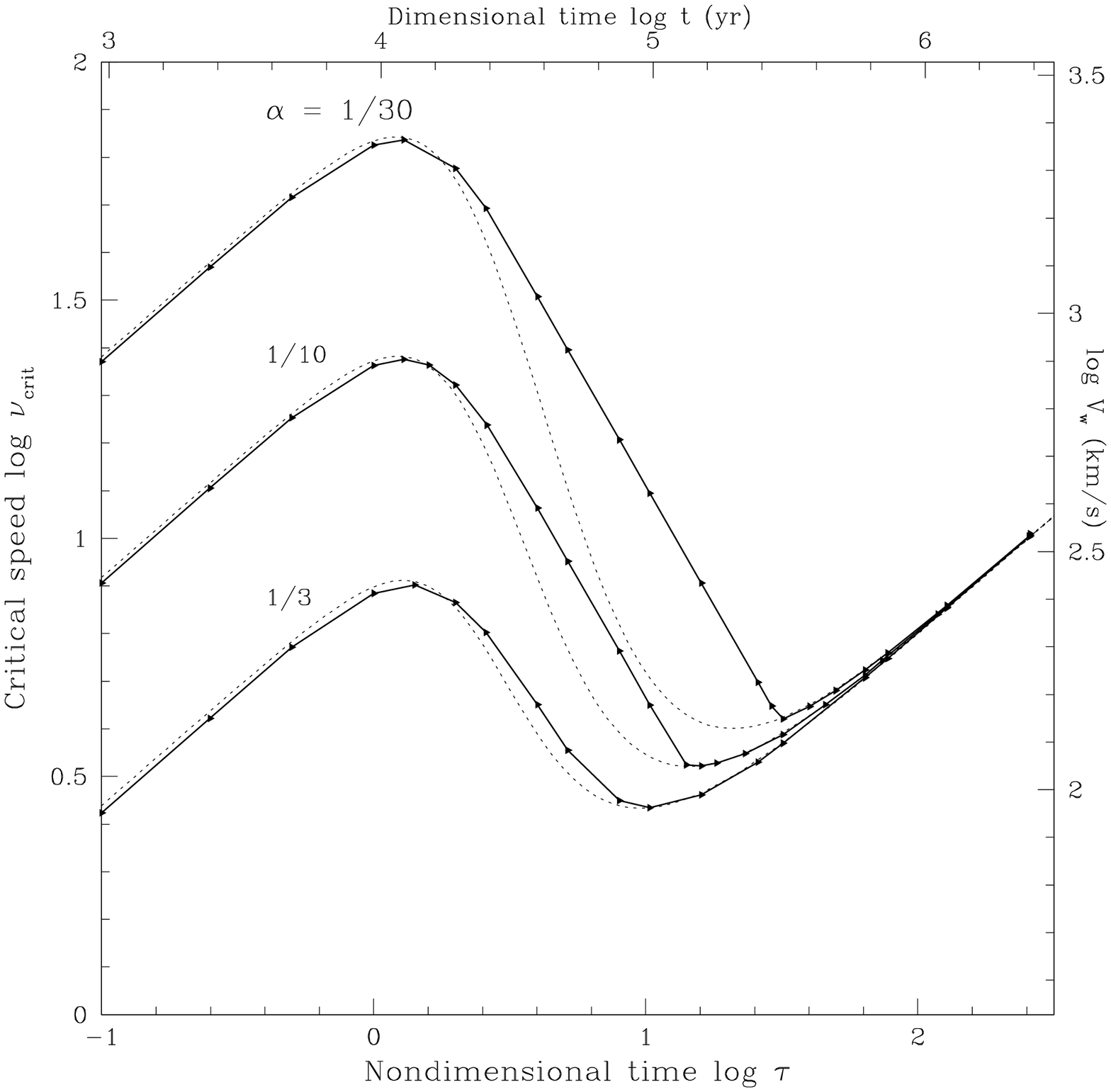}
\newpage
\plotone{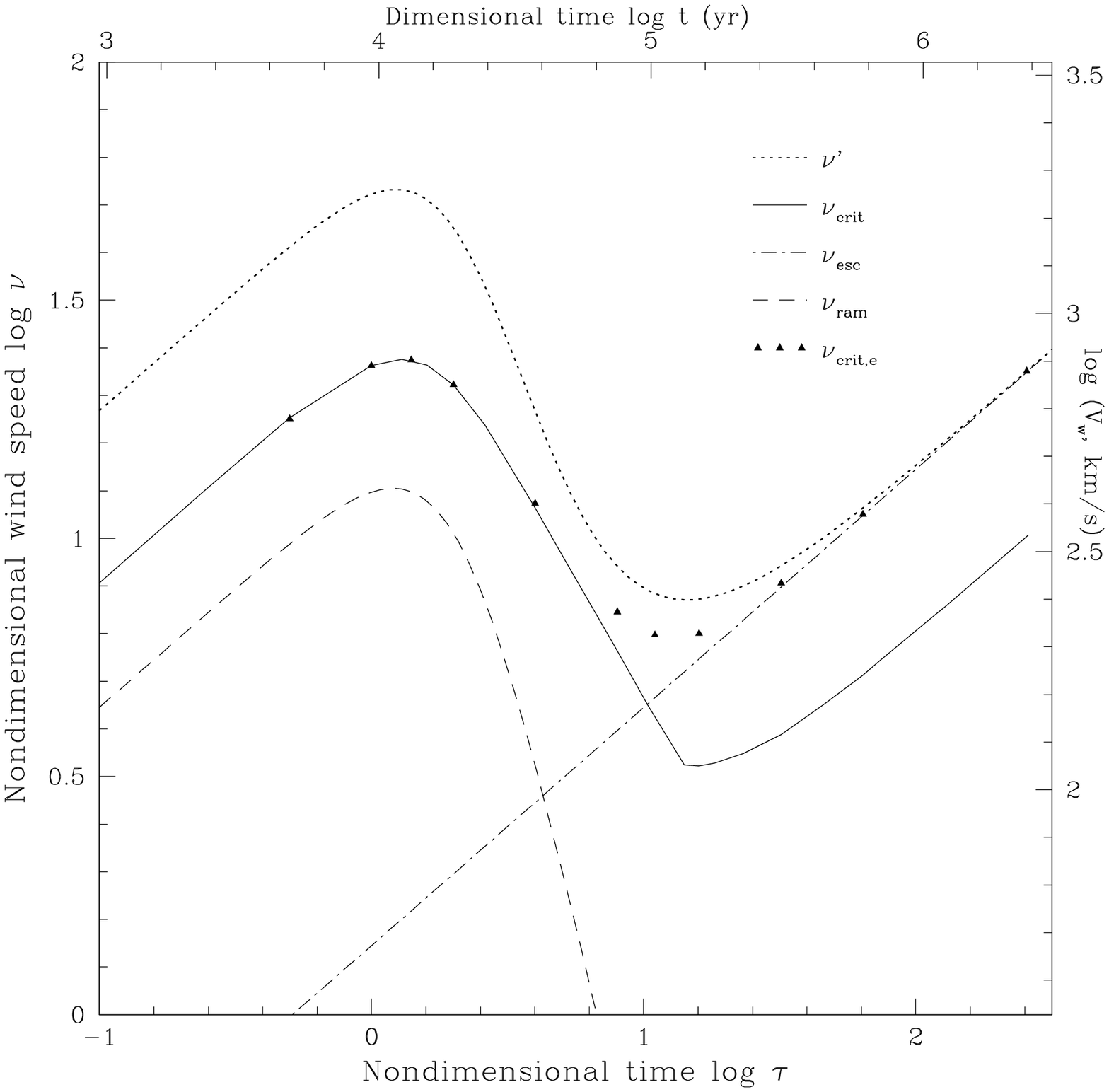}
\newpage
\plotone{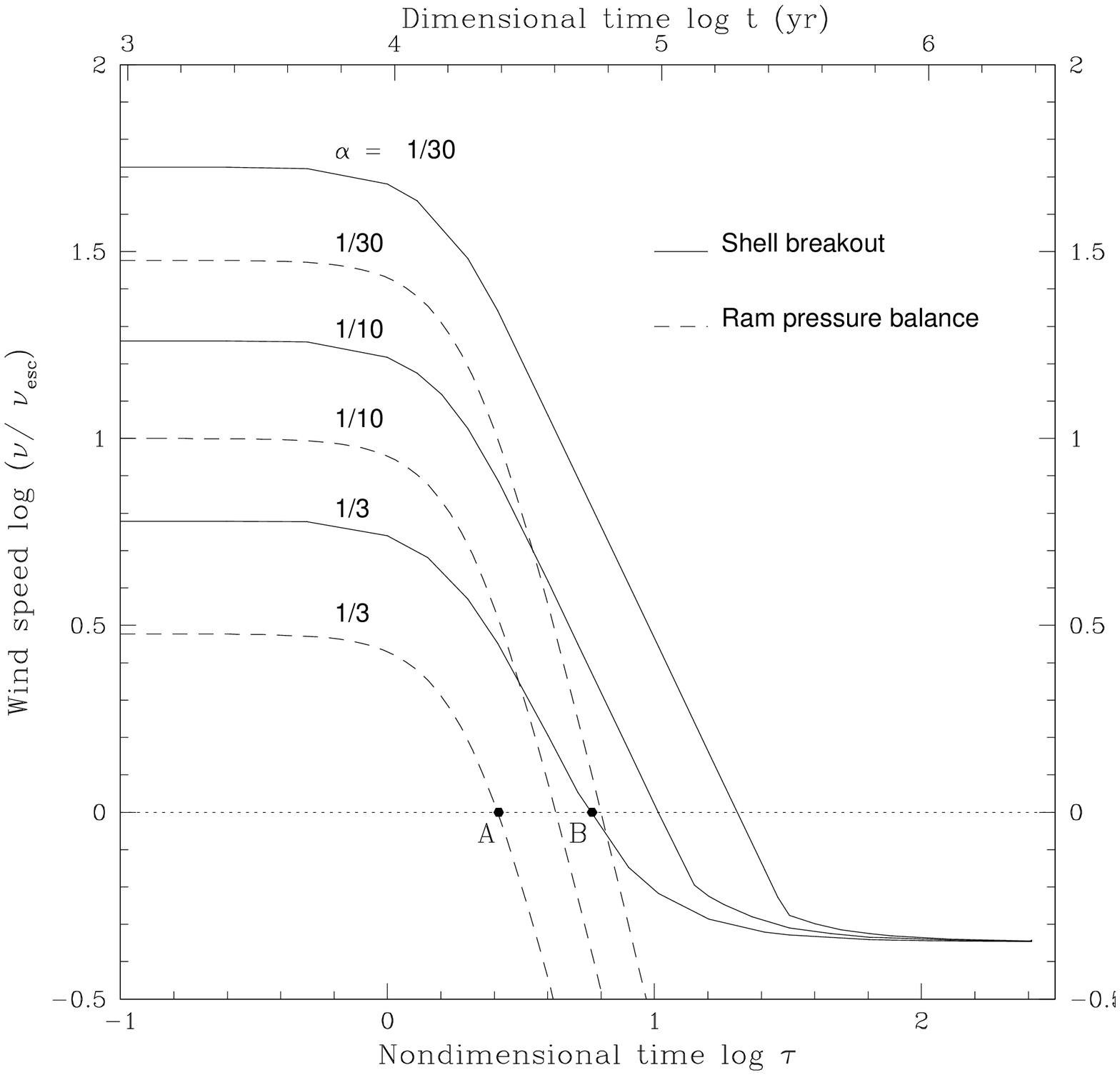}

\end{document}